\documentclass[lettersize,journal]{IEEEtran}
\usepackage{array}
\usepackage[caption=false,font=normalsize,labelfont=sf,textfont=sf]{subfig}
\usepackage{stfloats}
\usepackage{url}
\usepackage{verbatim}
\usepackage{cite}
\usepackage{amsmath,amssymb,amsfonts}
\usepackage{algorithm}  
\usepackage{algpseudocode}
\usepackage{graphicx}
\usepackage{textcomp}
\usepackage{xcolor}
\usepackage{booktabs}
\begin{document}

\title{Differentiated Federated Reinforcement Learning Based Traffic Offloading on Space-Air-Ground Integrated Networks}

\author{Yeguang Qin,~\IEEEmembership{Student Member, IEEE}, Yilin Yang,~\IEEEmembership{Student Member, IEEE}, Fengxiao Tang,~\IEEEmembership{Member, IEEE}, Xin Yao,~\IEEEmembership{Member, IEEE}, Ming Zhao,~\IEEEmembership{Member, IEEE}, and Nei Kato,~\IEEEmembership{Fellow, IEEE}

\thanks{This work was supported by the National Key R\&D Program of China (Grant no.2021ZD0140301), National Key R\&D Program of China (Project no.2020AAA0109602), Changsha Municipal Natural Science Foundation (Grant no.kq2208284), Hunan Provincial Natural Science Foundation (Grant no.2023jj40774), National Natural Science Foundation of China (Grant no.62302527), and the High Performance Computing Center of Central South University. (Corresponding author: Fengxiao Tang.)}

\thanks{Y. Qin, Y. Yang, F. Tang, X. Yao, and M. Zhao are with the School of Computer Science and Engineering, Central South University, ChangSha, China. Emails:\{qinyeguang, yangyilin, tangfengxiao, xinyao, meanzhao\}@csu.edu.cn }
\thanks{N. Kato is with Graduate School of Information Sciences (GSIS), Tohoku University, Sendai, Japan. Email:kato@it.is.tohoku.ac.jp}}



\maketitle

\begin{abstract}
The Space-Air-Ground Integrated Network (SAGIN) plays a pivotal role as a comprehensive foundational network communication infrastructure, presenting opportunities for highly efficient global data transmission. Nonetheless, given SAGIN's unique characteristics as a dynamically heterogeneous network, conventional network optimization methodologies encounter challenges in satisfying the stringent requirements for network latency and stability inherent to data transmission within this network environment. Therefore, this paper proposes the use of differentiated federated reinforcement learning (DFRL) to solve the traffic offloading problem in SAGIN, i.e., using multiple agents to generate differentiated traffic offloading policies. Considering the differentiated characteristics of each region of SAGIN, DFRL models the traffic offloading policy optimization process as the process of solving the Decentralized Partially Observable Markov Decision Process (DEC-POMDP) problem. The paper proposes a novel Differentiated Federated Soft Actor-Critic (DFSAC) algorithm to solve the problem. The DFSAC algorithm takes the network packet delay as the joint reward value and introduces the global trend model as the joint target action-value function of each agent to guide the update of each agent's policy. The simulation results demonstrate that the traffic offloading policy based on the DFSAC algorithm achieves better performance in terms of network throughput, packet loss rate, and packet delay compared to the traditional federated reinforcement learning approach and other baseline approaches.
\end{abstract}


\section{Introduction}
\IEEEPARstart{T}{he} optimized decision-making is a critical challenge in the Cyber-physical system, especially in the next-generation network of B5G/6G with a highly dynamic and large-scale environment. Reinforcement learning (RL) is one kind of machine learning technology to optimizes decision-making by maximizing the cumulative reward by continuously exploring and exploiting the environment of an agent. Using RL to obtain network optimization strategies is a current research hotspot in the field of modern networks, e.g., network resources allocation~\cite{xiong2020resource, wang2021} and traffic control~\cite{dong2021intelligent}.\par

However, traditional RL applications still suffer from critical problems in scenarios with complex environments. For example, the local strategy trap when the network environment becomes dynamic and heterogeneous. The local strategy trap arises when decisions are derived from incomplete local information, overlooking the broader implications of global environmental change. This leads to ill-informed outcomes due to a lack of holistic consideration. Due to the features of high dynamic and heterogeneous complex networks, the local strategy trap occurs frequently in network scenarios such as Internet of Things (IoT)~\cite{tang2019probe, guo2018}, vehicular network~\cite{tang2021comprehensive, zhang2020} and Space-Air-Ground Integrated Network (SAGIN)~\cite{tang2021deep, 9403380}. \par

Federated learning (FL) is an advanced learning technology to train a shared learning model without raw training data by considering the privacy of distributed devices in complex environments. 
By combining both RL and FL, a privacy-preserving multi-agent collaboration approach referred to as federated reinforcement learning (FRL) is proposed. 
In FRL, each agent trains the data separately, aggregates it into a uniform global policy model, and then distributes it to the agents, ensuring the user's data privacy as much as possible while integrating local learning results from each agent.\par

The local strategy traps are solved to some extent by the shared training manner of FRL as the global information is implicitly shared during the integrated learning process. 
Unfortunately, the global sharing-based FRL employing a global learning model for local inference may still fall into the local strategy trap. 
The accuracy of using the global learning model for local inference is ensured by the assumption that the environments of diverse devices are homologous with the same state transition probability, which is not practical as the considered state transition probability is always differentiation in the heterogeneous environment.\par

For example, a dynamic and heterogeneous SAGIN consists of satellites, unmanned aerial vehicles (UAVs), ground-based stations in different regions, and user devices. 
In such a network environment, there is significant dynamism and heterogeneity among regions due to the high-speed movement of nodes, the extensive range of data transmission, and the convergence between different types of networks\cite{ASurvey}. 
The differentiation between regions is generally due to network state differences, sample distribution differences, and dynamic characteristic differences. 
These discrepancies lead to the global policy model obtained by the aggregation of agents in each region can only be suboptimal compared to the optimal policy in each region. 
It is not easy to use the traditional FRL approach to obtain a global policy model to address issues like traffic offloading in SAGIN that performs well across all differentiated regions.\par

\begin{figure}[t]
	\centering
	\includegraphics[width=\columnwidth]{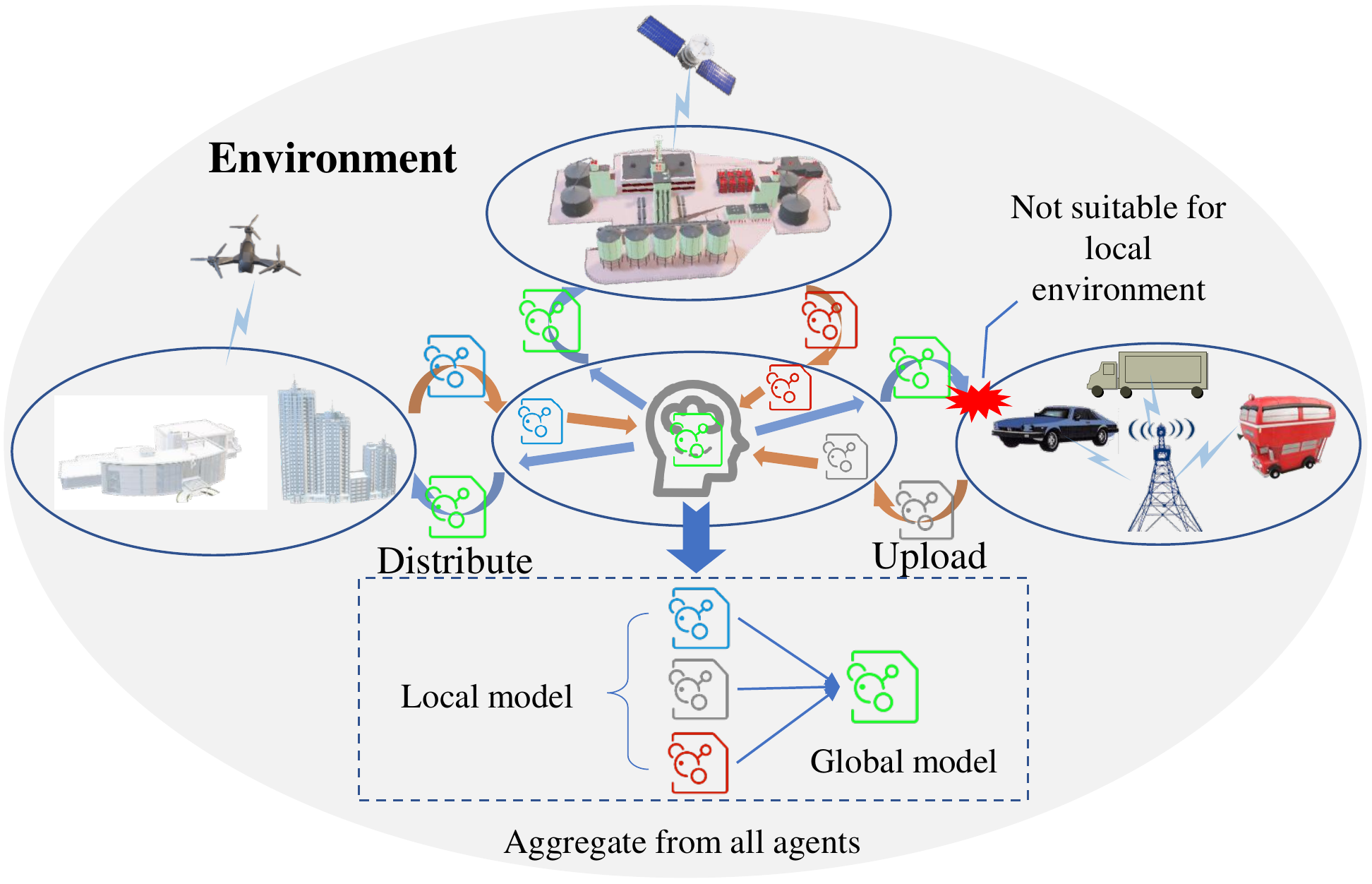}
	\caption{The global learning model does not work well because of environmental differentiations.}
	\label{fig:DFRL}
\end{figure}

As shown in Fig.~\ref{fig:DFRL}, traditional FRL distributes a uniform global learning model to each region. 
When there are differentiations in the environment of each region, the decisions inferred by the global learning model may be misjudged. 
Therefore, considering the differentiations between regions in complex network environments, we propose a novel differentiated federated reinforcement learning (DFRL) approach. 
Instead of seeking a uniform global policy model, DFRL enables agents across regions to cooperate in training their respective policy models.

Precisely, the cooperation among the agents in each region is accomplished through the global trend model. 
The global trend model is generated by aggregating the trend model developed by all the agents involved in the cooperation, reflecting the state of the environment and its changing trend in each region. 
Then, under the guidance of the global trend model, the agents in each region get their local policy models by training with local data. The introduction of the global trend model allows agents to consider each region's differentiations based on the information from multiple regions to obtain local strategies that are more applicable to their regions.\par 

The main work of the paper can be summarized as follows:
\begin{itemize}
	\item In this paper, we study the traffic offloading problem in a dynamic heterogeneous network like SAGIN and model it as a DEC-POMDP decision problem with the objective of optimizing the network delay.
  \item Then, we propose a novel concept of differentiated federal reinforcement learning by introducing a trend model, and based on this concept, an algorithm named Differentiated Federated Soft Actor-Critic (DFSAC) is proposed to solve the joint policy in DEC-POMDP.
  \item We design a SAGIN struct and propose a dynamic traffic offloading method based on the DFSAC algorithm to solve the traffic offloading problem under this complex network structure. Simulation results show that our method achieves better results regarding network throughput, packet loss rate, and delay than other methods.
\end{itemize}

The rest of the paper is organized as follows. In section \uppercase\expandafter{\romannumeral2}, we discuss related work. In section \uppercase\expandafter{\romannumeral3}, we detail the system model and the formulation of the problem. Section \uppercase\expandafter{\romannumeral4}  elaborates on the concept of DFRL and proposes the DFSAC algorithm.
An empirical study of traffic offloading in SAGIN is presented in Section \uppercase\expandafter{\romannumeral5} and summarized in Section \uppercase\expandafter{\romannumeral6}.

\section{Related Work}
\subsection{Federated Learning}
FL is a distributed machine learning approach in privacy-preserving scenarios, where the critical point is that information is passed between collaborators by sharing models rather than data~\cite{konevcny2016federated}.
In recent years, researchers have proposed several solutions to the challenges faced in FL.
For example, to obtain higher communication efficiency, McMahan et al. proposed FedAVG~\cite{mcmahan2017communication}, which is now widely used as the baseline for FL research. FedAVG is a widely adopted federated learning algorithm that aggregates model parameters using weighted averaging by uploading local model parameters to a central server, computing the average of all model parameters, and then broadcasting this average to all local devices.
To address the heterogeneity problem in FL, Yuan, Dinh, and Ruan et al. proposed~\cite{yuan2021federated},~\cite{t2020personalized}, and~\cite{ruan2021towards}, respectively, to solve data heterogeneity, model heterogeneity, and device heterogeneity.
Moreover, Qu et al. further analyzed the convergence of FL methods in ~\cite{qu2020federated}, comprehensively studying how the convergence of FedAVG varies with the number of participating devices in the FL setting. They demonstrate the convergence of the FedAVG method in several differentiated scenarios and perform a comprehensive study of its convergence rate.
This work provides a solid foundation for federated learning in further research.

\subsection{Federated Reinforcement Learning}
RL is a branch of machine learning (ML).
Compared with other machine learning methods such as supervised or unsupervised learning methods, RL generates samples and learns through these samples through constant interaction between the agent and the environment~\cite{kaelbling1996reinforcement}.
The RL is widely used for communication and networking optimization in various networks. Wang et al.~\cite{wang2022ABRLA} proposed a semantic communication framework for efficient textual data transmission in resource-constrained wireless networks. This framework utilizes knowledge graphs and a proximal-policy-optimization-based RL algorithm integrated with an attention network to optimize the partial transmission of semantic information and enhance semantic similarity metrics.
 Luan et al.~\cite{luan2021epc} employed deep reinforcement learning (DRL) to ascertain the optimal traffic allocation ratio among multiple controllable paths for source-destination pairs. Gao et al.~\cite{gao2022iotj} proposed a deep reinforcement learning-based framework for joint optimization of computing, pushing, and caching in mobile edge computing networks, effectively reducing transmission bandwidth and computing cost through dynamic orchestration and proactive content delivery.
However, when the environment gets bigger, it becomes difficult for a single agent to complete the complex task in these scenarios due to constraints such as information processing capabilities or learning efficiency.
The natural idea is to set up multiple agents in the environment and complete the task by cooperating among them.
Hence, researchers have introduced distributed RL and parallel RL ~\cite{grounds2005parallel} with the aim of accelerating the learning of optimal strategies for single-agent RL problems through parallel hardware utilization. However, these approaches also raise concerns about potential agent privacy breaches.

In this context, the integration of FL, a paradigm centered on preserving privacy, into the realm of reinforcement RL represents a notable development. FRL is a novel, distributed, and collaborative methodology that effectively amalgamates the principles of FL and RL. In this paradigm, each agent trains data locally and builds a shared model, which protects agent privacy and accelerates agent learning efficiency~\cite{qi2021federated}.\par

\subsection{Federated Learning in Heterogeneous Network}

Environmental heterogeneity is a significant challenge for FL in real application scenarios. For this challenge, Yuan et al. proposed an improved Federated Deep AUC (area under the ROC curve) Maximization algorithm in~\cite{yuan2021federated} to solve the data heterogeneity problem, while Hanzely et al. proposed a hybrid global model and local model to achieve balanced training in ~\cite{hanzely2020federated} to solve the conflicts caused by data heterogeneity. Shen et al. proposed a novel distributed learning scheme~\cite{shen2023} that combines FL with a model-splitting mechanism to accommodate customer heterogeneity.
In addition, researchers have also tried to use personalized federated learning methods~\cite{zhang2022personalized, shamsian2021personalized, deng2020adaptive} to address the challenges caused by environment heterogeneity.\par

In communication networks, Liu et al. proposed a method for Radio Access Network (RAN) slicing using the FRL approach in~\cite{liu2020device}, which improves the communication efficiency and throughput of the network.
Kwon et al. proposed an FRL-based resource allocation method for Internet of Things Underwater (IoUT) devices in~\cite{kwon2020multiagent}, which has significant advantages over the single-agent DRL approach.
Zhu et al. proposed a fast convergent federated-based dynamic task offloading method in the power grid Internet of Things~\cite{9449265} but needed to sufficiently consider the differentiation of environments.
Wang et al.~\cite{wang2023} combined mobile edge computing with fiber-wireless networks and applied a reputation-based FRL strategy, effectively optimizing network performance and resource allocation in IoT deployments while protecting user privacy.\par

The current methods for addressing environmental heterogeneity primarily involve various forms of aggregation or training of local models. However, these methods lack the capability to dynamically adapt to environmental changes. As large-scale network environments become increasingly dynamic and heterogeneous, these methods struggle to effectively adapt to variations in regional network conditions and network types. This leads to reduced learning efficiency and potential convergence issues, posing a significant challenge.\par

\subsection{Offloading Solutions in SAGIN}
Guo et al.~\cite{guo2020} have devised a comprehensive energy-efficient optimization strategy that integrates UAV-assisted communication with Mobile Edge Computing (MEC), effectively reducing overall energy consumption through strategic task offloading, transmission bit allocation, and UAV trajectory planning. Li et al.~\cite{Li2020} proposed an integrated satellite/terrestrial collaborative transmission scheme, incorporating cache-enabled Low Earth Orbit (LEO) satellites as part of the RAN, which offloads traffic from base stations through satellite broadcast transmission, thereby achieving energy-efficient RAN operations.

The existing research predominantly addresses scenarios involving either satellites or UAVs in isolation; however, given the inherent complexity, heterogeneity, and dynamic nature of SAGIN, these systems present considerable challenges in terms of accurate modeling. As a result, the model-free methodology of reinforcement learning has been extensively investigated and applied to effectively tackle the offloading issues within the SAGIN framework.
Cheng et al.~\cite{cheng2019} have developed an innovative SAGIN edge/cloud computing architecture, tailored for offloading computation-intensive applications while considering remote energy and computational constraints. Tang et al.~\cite{tang2021deep} have presented a reinforcement learning-based traffic offloading scheme, taking into account the substantial node mobility within SAGIN, as well as the frequent variations in network traffic and link statuses. 
Zhang et al.~\cite{zhang2023} proposed a Learning-based Orbital Edge Offloading (LOEF) method using multi-agent learning, enabling UAVs to coordinate and learn optimal offloading strategies for computational task scheduling in the Internet of Remote Things (IoRT) within SAGIN.

Although existing solutions have made significant progress in SAGIN, they often overlook the critical aspect of privacy protection. In this context, FRL offers a potential mechanism for privacy preservation. However, the application of FRL faces inherent limitations, mainly because it relies on a unified global policy model to guide the reasoning process of local agents. This approach can lead to suboptimal results in local reasoning, thereby affecting the quality of overall decision-making. Therefore, applying FRL in dynamic and heterogeneous SAGIN environments to effectively solve the traffic offloading problem remains a significant challenge.

\section{SYSTEM MODEL}
\subsection{Network Model}

\begin{figure}[h]
  \centering
  \includegraphics[width=\linewidth]{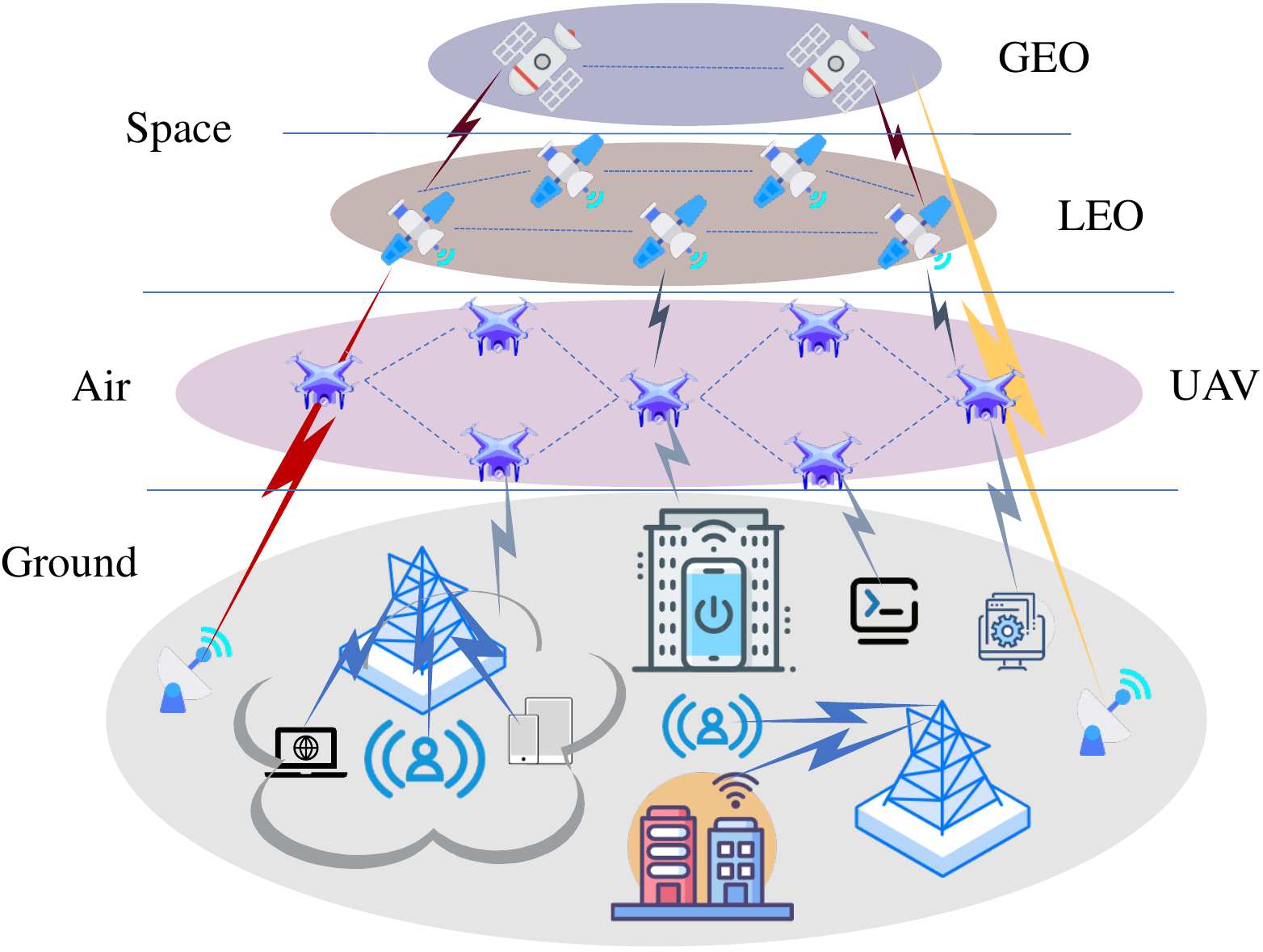}
  \caption{The four-layer SAGIN architecture.}
  \label{fig:SAGIN}
\end{figure}

In this paper, we consider traffic offloading in SAGIN and establish a multi-dimensional heterogeneous network model: The ground network is composed of mobile user equipment (UE), base stations (BS) and edge base stations. The air network includes a series of dynamically deployable UAVs as an extension of the ground network. The space network is a double-layer communication satellite network composed of LEO and Geostationary Earth Orbit satellites (GEO). The network structure is shown in Fig.~\ref{fig:SAGIN}.\par 

Due to the uneven distribution of UE in the ground network, the traffic may be over the load of the BS in service-intensive areas. Therefore, we use UAV as a mobile flying base station to dynamically relay the offloaded traffic to other regions to avoid network congestion. We set the trajectory of the UAV as circular motion or random movement. When the UAV flies to an area, it can be used as the traffic relay node of the area. In this paper, the trajectory optimization of UAV in SAGIN is not our focus, so we do not introduce it additionally. However, the signal coverage of UAV is limited. In the process of moving, UAV may fly out of the target area, resulting in signal loss and a sharp increase in the packet loss rate. On the other hand, the loadable traffic of UAV is relatively limited, and a large number of dropped packets exceeding the capacity will also cause an increase in packet loss rate. Therefore, the communication satellite with large coverage is introduced as the available traffic relay node, and the ground traffic can be offloaded to available UAVs or communication satellites. Our network model further considers the double-layer satellite network composed of LEO and GEO based on the existing research on traffic offloading in SAGIN. The two-layer satellite network expands the available resource pool of SAGIN, and the use of GEO also improves the stability of the network and the overall performance.\par

In this paper, we model a four-layer SAGIN as an undirected graph ${G=\left( V,E \right)}$. Among them, ${V=\left\{ {{V}^{D}},{{V}^{B}},{{V}^{U}},{{V}^{L}},{{V}^{G}} \right\}}$ represents different types of nodes in SAGIN. ${{V}^{D}}=\left\{ v_{1}^{D},v_{2}^{D},\cdots ,v_{n}^{D} \right\}$ is a collection of UE nodes, $n$ is the number of this type nodes. Similarly, ${{V}^{B}}$ is the collection of BS nodes, ${{V}^{U}}$ is the collection of UAV nodes, ${{V}^{L}}$ and ${{V}^{G}}$ are collections of LEO nodes and GEO nodes. Each node in ${V}$ contains its attribute parameters and network state parameters. When the node encounters buffer overflow or connection breakage during packet transmission, it will discard these packets. And the collection of edges $E=\left\{ e_{11}^{DB},e_{12}^{DB},\cdots,e_{ij}^{xy} \right\}$ indicates the link between any two nodes in the network. For example, $e_{ij}^{xy}$ is the link between node $v_{i}^{x}$ and $v_{j}^{y}$, among them $v_{i}^{x},v_{j}^{y}\in V$. The model will dynamically change the value of $E$ according to the location and connection state of each node in the network. Further, ${{V}^{D}}$ is divided into two types: source node and target node. The source node is responsible for sending data packets to the target node according to the preset generation mode. Other type nodes in $V$ are relay nodes used to transfer data packets from the source node to the target node. Among them, base station nodes ${{V}^{B}}$ will make the offloading decision as the traffic offloading node.\par

\subsection{Transmission Model}

Since SAGIN consists of several different types of nodes, the transmission model between the nodes is not the same. For the transmission between UAV and BS, according to \cite{shi2019multi}, its path loss can be found using the following equation:
\begin{equation}\label{(1)}
  PL\left(l_{U\!B}, \omega\right)=10 \varphi \log \left(l_{U\!B}\right)+\eta\left(\omega-\omega_{0}\right) e^{\frac{\omega_{0}-\omega}{\gamma}}+k_{0}
\end{equation}
where $l_{U\!B}$ denotes the horizontal distance between UAV and BS, $\omega$ is vertical angle between UAV and BS, $\omega_{0}$ is the angle offset. And $\varphi$, $\eta$ and $k_{0}$ are the terrestrial path loss exponent, excess path loss and excess path loss offset, $\gamma$ is the angle scalar. Due to the dynamic nature of the UAV nodes, the distance between UAV and BS changes over time. We use the model in \cite{lyu2017optimal} to calculate the transmission rate between UAV and BS during time slot t, with the transmission rate as follows:
\begin{equation}\label{(2)}
  \nu_{t}^{U\!B}=B_{U\!B} \log _{2}\left(1+\frac{P_{U\!B} \cdot 10^{-\frac{PL}{10}}}{\sigma_{U\!B}^{2}}\right)
\end{equation}
where $B_{U\!B}$ is the channel bandwidth between UAV and BS, $P_{U\!B}$ represents the transmission power, $PL$ represents the path loss, and $\sigma_{U\!B}^2$ represents the power of background noise.\par

In our transmission model, the distance between the satellite and UAV is much greater than the movement range of UAV and UAV height. Thus, UAVs can be considered ground devices along with the base station during satellite transmission. We consider the rain attenuation during satellite-to-ground communication as a Weibull-based stochastic process \cite{kanellopoulos2014channel}. Therefore, the channel gain from the UAV to the satellite can be calculated by the following equation:
\begin{equation}\label{(4)}
I_{U\!S}=\frac{G_{U\!S} G_{SU} \lambda_{US}^{2}}{\left(4 \pi l_{U\!S}\right)^{2}} 10^{-\frac{F_{rain}}{10}}
\end{equation}
where $G_{U\!S}$ and $G_{SU}$ are the antenna gains of the UAV and the satellite, respectively, $\lambda_{U\!S}$ is the wavelength, and $l_{U\!S}$ is the distance between the satellite and the UAV. Rain attenuation $F_{rain}$ is modeled by Weibull distribution \cite{kanellopoulos2014channel}. Similarly, the channel gain from the BS to the satellite can be calculated as follows:
\begin{equation}\label{(5)}
I_{B\!S}=\frac{G_{B\!S} G_{S\!B} \lambda_{B\!S}^{2}}{\left(4 \pi l_{B\!S}\right)^{2}} 10^{-\frac{F_{rain}}{10}}
\end{equation}
where $G_{B\!S}$ and $G_{S\!B}$ are the antenna gains of the BS and the satellite respectively, $\lambda_{B\!S}$ is the wavelength, and $l_{B\!S}$ is the distance between the BS and the satellite. After getting the channel gain, we can calculate the transmission rate between the UAV and the satellite as follows:
\begin{equation}\label{(6)}
\nu_{t}^{U\!S}=B_{U\!S} \log _{2}\left(1+\frac{P_{U\!S} \cdot\left|I_{U\!S}\right|}{\sigma_{U\!S}^{2}}\right)
\end{equation}
where $B_{U\!S}$ represents the channel bandwidth of the UAV and the satellite communication link, $P_{U\!S}$  represents the transmission power between the UAV and the satellite, $I_{U\!S}$  represents the channel gain between the UAV and the satellite, and $\sigma_{U\!S}^{2}$ is the power of the background noise. Additionally, the transmission rate between the BS and the satellite can be obtained as follows:
\begin{equation}\label{(8)}
\nu_{t}^{B\!S}=B_{B\!S} \log _{2}\left(1+\frac{P_{B\!S} \cdot\left|I_{B\!S}\right|}{\sigma_{B\!S}^{2}}\right)
\end{equation}
where $B_{B\!S}$ denotes the channel bandwidth between the BS and satellite communication link, $P_{B\!S}$ denotes the transmission power between the BS and satellite, $I_{B\!S}$ is the channel gain between the BS and satellite, and $\sigma_{B\!S}^{2}$ is the power of the background noise.\par

In addition, the transmission between GEO and LEO adopts the free space propagation model. We can calculate the channel gain of inter-satellite communication as follows:
\begin{equation}\label{(satelliteGain)}
	I_{H\!L}=\frac{G_{H\!L} G_{LH} \lambda_{HL}^{2}}{\left(4 \pi d_{H\!L}\right)^{2}}
\end{equation}
where $G_{H\!L}$ and $G_{LH}$ are the antenna gains of GEO and LEO, $\lambda_{H\!L}$ is the wavelength, and $d_{H\!L}$ is the distance between the GEO and LEO. The transmission rate of GEO and LEO is as follows:
\begin{equation}\label{(satelliteRate)}
	\nu_{t}^{H\!L}=B_{H\!L} \log _{2}\left(1+\frac{P_{H\!L} \cdot\left|I_{H\!L}\right|}{\sigma_{H\!L}^{2}}\right)
\end{equation}
where $B_{H\!L}$ is the channel bandwidth between the GEO and LEO communication link, $P_{H\!L}$ is the transmission power between the GEO and LEO, $I_{H\!L}$ denotes the channel gain between the GEO and LEO, and $\sigma_{H\!L}^{2}$ denotes the power of the background noise.\par
\subsection{Problem Formulation}

Denoted by $X_t$, the total amount of data packets in $t$ th time slot. $X_t$ follows a normal distribution and these packets are transmitted to node $v_d\in V$ through the set routing path ${{L_t}=\{v_x, v_y, \ldots, v_d\}}$. This process can be expressed as ${M}_{v_{s}\rightarrow {v}_{d}}=\left\{{X}_{t}, {~L}_{t}, {~T}, ~r\right\}$, $X_t$ is the amount of data, $L_t$ is the defined routing path, $T$ is the transmission delay of the packet, and $r$ is the transmission rate of the packet. The system will generate a new routing path when using the traffic offloading algorithm $A(m)$ during data packet transmission. $m$ represents the process of transmitting this data packet. At this time, there will be UAV or satellite nodes in the path. In this paper, our goal is to minimize the packet delay of the entire network over the designed time $D$ by the usage of the traffic offloading method to generate a new routing path at time $t$ dependent on specific parameters of:


\begin{equation}\label{eq:10}
Z_{\mu}(m, t) = \left\{\begin{aligned}
&m_{X_t, L_t, T,r}, && a=0 \\
&A\left(m_{X_t, L_t, T,r}\right),  && a \neq 0
\end{aligned}\right.
\end{equation}
where $a = 0$ means the offloading method is not used, and $a \neq 0$ means that the offloading method is used.

\begin{equation}\label{(11)}
\min _{\mu} \sum_{D} \sum_{m} O\left(Z_{\mu}(m, t)\right)
\end{equation}
where $\mathrm{O}(\cdot)$ calculates the delay of all packets in the entire network. We are committed to minimizing the delay of data packets by minimizing the parameters of the traffic offloading method.\par

\section{Traffic Offloading Method}

To address the above issues, we define SAGIN as a differentiated environment consisting of a series of dynamically heterogeneous local environments. Since the steady-state Markov process is not applicable to this environment, this study will describe the process using the framework of the DEC-POMDP\cite{bernstein2002complexity}, which is an extended form of the partially observable Markov decision process. ``Distributed'' implies that the training of each intelligence in the process is decentralized. ``Partially observable'' implies that each agent can only observe a part of the environment. Therefore, this study formalizes the local policy trap problem in SAGIN as an environment discretization problem under DEC-POMDP to obtain the joint optimal traffic offloading policy in such a differentiated environment.
Among them, the BS located in each region are considered agents.
During packet transmission, BS will use the traffic offload algorithm $A$ to make traffic offload decisions.
And, the traffic offloading problem in SAGIN is transformed into DEC-POMDP and thus trained to learn.

In DEC-POMDP, multiple agents need to work together to maximize the overall reward in a partially observable environment, where each agent can only partially observe the environment state, and cooperation among the agents is required to solve the task. DEC-POMDP represents the connections among the agents through an interrelationship model between the agents and takes action based on this. The DEC-POMDP model can be defined as:
\begin{equation}\label{(DECPOMDP)}
  M = \langle S, A_1, ..., A_N, T, R, \Omega_1, ..., \Omega_N, O, \gamma \rangle
\end{equation}
among them:
\begin{itemize}
  \item $S$ is the state space, representing all possible states of the environment. In the traffic offloading problem discussed in this study, this state space can be considered as the network state information of all network nodes which includes both dynamically changing data and covers static characteristics such as generalized mobility model, service generation rate, and queue maximum size. This is not complete for a single agent to make observations. 
  \item $A_i=\left\{0, op_j\right\}$ denotes the action space of agent $i$. Here, the first element signifies that the system forwards the packet to the next node following the pre-set routing path. The second element, the offloading path $op_j$, involves the selection of the next relay node, such as UAV, LEO, or GEO, to determine the new transmission path for the packet.
  \item $T: S \times A_1 \times ... \times A_N \times S \rightarrow [0,1]$ is a state transfer function that represents the probability distribution of transfer to a new state given the actions of all agents and the current state of the environment. Since it is difficult to build a corresponding mathematical model for the state transfer probability of the environment, a model-free approach will be used to represent it. i.e., the training is guided by neural network models to predict future reward sums.
  \item $R: S \times A_1 \times ... \times A_N \rightarrow \mathbb{R}$ is reward function. Our optimization goal is to minimize the total delay of the network, so set the reward function to
  \begin{equation}\label{(14)}
  R(s_{t},a_{t})=\left\{
    \begin{split}
    & \frac{1}{D_{t}}, & arrive \\
    & -(T_{drop}-T_{born}), & drop
    \end{split}
  \right.
  \end{equation}
  where $D_t$ denotes packet delay, $T_{drop}$ and $T_{born}$ are the packet drop time and born time. When the packet arrives at the destination node, a positive reward is given according to the delay, and the lower the delay, the higher the reward value. If the packet is dropped, a negative reward is given according to the difference between the drop and the birth time.
  \item $\Omega_i$ denotes the observation space of agent $i$, which is defined as a multi-dimensional array encompassing all relevant information of the one-hop and two-hop neighboring nodes in its vicinity. The observational outcomes of the agent are constituted by the network information of these one-hop and two-hop neighbors, meticulously gathered through the use of the Hello protocol. For any given time point $t$, the local observation space $\Omega_i$ for the agent can be defined as:
  \begin{equation}\label{(jubu)}
  \Omega_{i}(t) = \left\{ \mathcal{Q}_{t}^{adj(i)}, \mathcal{Q}_{t}^{adj(adj(i))} \right\}
\end{equation}
    where $\mathcal{Q}_{t}^{adj(i)}$ represents the collection of relevant information at time 
$t$ for the immediate neighbors of the agent 
$i$. $\mathcal{Q}_{t}^{adj(adj(i))}$represents the collection of relevant information at time $t$ for the neighbors of the neighbors of agent $i$. $adj(i)$ denotes the set of adjacent nodes to agent $i$.
  \item $O: S \times A_1 \times ... \times A_N \times \Omega_1 \times ... \times \Omega_N \rightarrow [0,1]$ is an observation probability function that represents the probability distribution of its local observations observed by agent $i$, given the actions of all agents and the current state of the environment.
  \item $\gamma \in (0, 1)$ denotes the discount factor used to measure the importance of immediate and future rewards.
\end{itemize}

In DEC-POMDP, the goal of the agents is to maximize the joint reward of all agents, i.e., the total expected discounted reward.
So, we can define the policy $\pi_i$ of agent $i$ as a mapping from the observation space $\omega_i$ to the action space $A_i$, indicating which action agent $i$ should take given its local observations. The joint policy $\Pi$ is then defined as the combination of all the agents' policies.
Thus, this paper aims to solve the joint traffic offloading policy $Pi$ for each agent in SAGIN.

\subsection{Differentiated Federated Reinforcement Learning}

\begin{figure}[h]
  \centering
  \includegraphics[width=\linewidth]{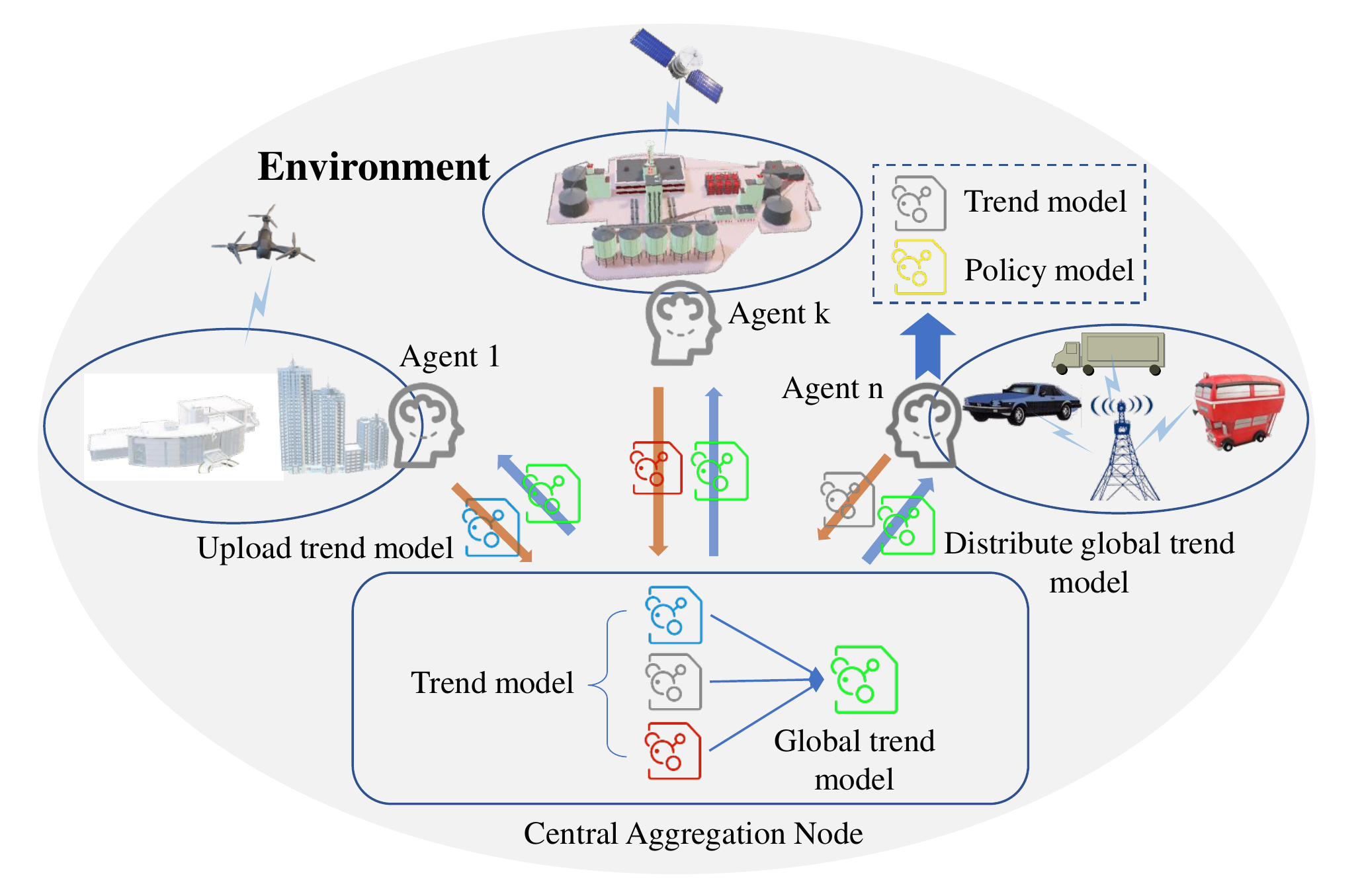}
  \caption{The agents communicate through the trend model.}
  \label{fig:TrendModel}
\end{figure}

In recent years, many scholars have proposed various algorithms to solve the joint policy in DEC-POMDP, including a value-based iterative approach, policy gradient-based approach, Monte Carlo tree search-based approach, and so on.
In this paper, the concept of DFRL is proposed to solve this joint policy $\Pi$ as shown in Fig.~\ref{fig:TrendModel}.\par 

The main improvement between DFRL and FRL is that DFRL isolates local learning from the global learning process. We first divide the local learning model of the agent into two parts: the trend model and the policy model. The policy model generates policies based on the local environment state, which is isolated and never shared with other agents.
The trend model is the window to communicate with the center node during the global learning process and guide the update of the policy model. 
And The global trend model is generated by the local trend model and shared among all agents, which is responsible for describing the state of the global environment, changing trends, and transferring the knowledge learned by all the agents. \par

The introduction of the trend model separates the learning of policy models, thus allowing agents to cooperate in obtaining policy models that satisfy their preferences. More importantly, the training process of the policy models of each agent is still distributed in the training nodes themselves, thus alleviating the environmental variability problem while satisfying the DEC-POMDP requirements, i.e., the non-IID data and local strategy trap problem in dynamically heterogeneous networks. Otherwise, separating the policy models ensures the privacy security of the policy models to meet the requirements of FL.\par
Theoretically, the DFRL architecture can be applied to many RL algorithms, such as Deep-Q network (DQN)~\cite{mnih2013playing} and Actor-critic methods~\cite{konda1999actor}. In the following section, we describe the concept of DFRL in the context of specific algorithms.

\subsection{Algorithms: Differentiated Federated Soft Actor-Critic}
In this section, we propose a novel DFRL algorithm called Differentiated Federated Soft Actor-Critic (DFSAC), which introduces the concept of DFRL based on the Soft Actor-Critic (SAC) algorithm\cite{christodoulou2019soft}.
Compared with the traditional FRL algorithm, this algorithm can better adapt to the differentiation between the environments in which the agents are located.\par

Due to data heterogeneity and node dynamics, the traditional FRL algorithm is not capable to handle the cooperated learning tasks in differentiated environments.
Therefore, we propose the DFSAC, which evolves SAC based on the concept of DFRL. 
The SAC algorithm is a DRL method consisting of Actor and Critic networks to optimize random policies in an off-policy manner.
The Actor network is used to output the actions, and the Critic network is used to evaluate the states and actions.
Its core feature is entropy regularization, where policy training trades off maximizing expected reward and entropy.
Increasing entropy makes the policy explore more, speeding up the subsequent learning process and preventing the policy from prematurely converging to a local optimum.
This makes it well-suited for exploring optimal strategies in several differentiated environments. Therefore, we choose this algorithm as our infrastructure.
In the DFRL framework, the Actor network is employed as the policy model, which is responsible for determining the actions to be taken based on the current state. On the other hand, the Critic network serves as the trend model, providing feedback on the actions chosen by the Actor network by estimating the value function. Within the Critic network, the target-critic component is specifically utilized as the global trend model, which plays a crucial role in the aggregation process by integrating local and global trend information to enhance the decision-making strategy in DFRL.
\begin{figure*}[t]
	\centering
 
	\includegraphics[width=0.8\linewidth]{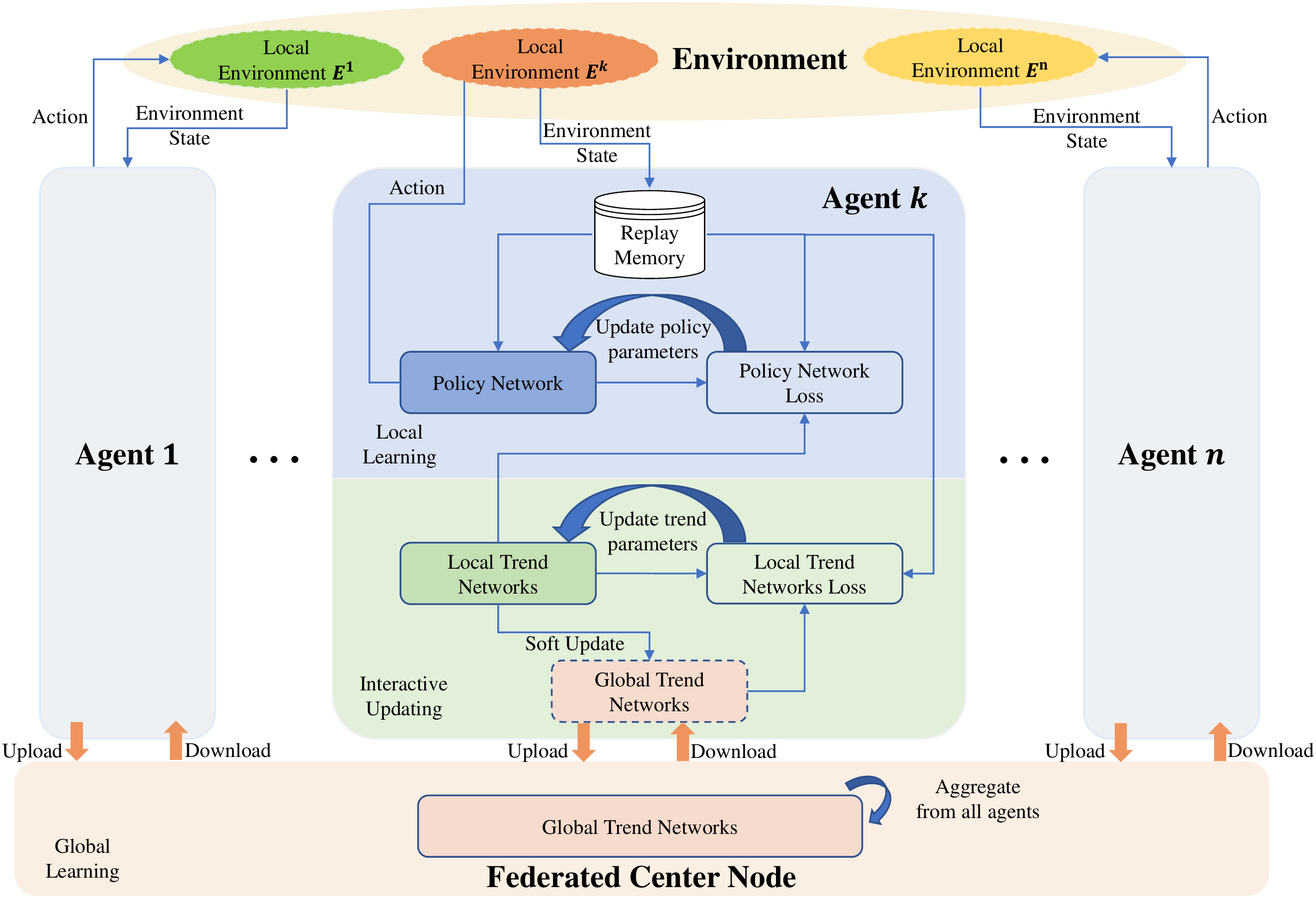}
	\caption{The learning process of DFSAC algorithm.}
	\label{fig:algorithm}
\end{figure*}
The DFSAC algorithm is processed by multiple agents and a federated center node, where the agents generate local trend networks and policy networks through the local environment, and the federated center node is responsible for collecting the local trend networks of each agent and aggregating them into global trend networks then distributing them to the agents.
Fig.~\ref{fig:algorithm} shows the whole learning process including global learning, interactive updating, and local learning, and the global trend model participates in all the processes in any agent $k$.
We define a global environment $E=\left\{ E^1,\cdots,E^k,\cdots,E^n \right\}$ consisting of a set of $N$ differentiated local environments.
A corresponding agent exists for each local environment, i.e. environment $E^k$ corresponds to agent $k$.
Unlike the traditional FRL approach, the goal of the DFSAC algorithm is to obtain a set of policies $\varPi$ applicable to each local environment $\varPi =\left\{ {{\widetilde{\pi }}^{1}},\cdots,{{\widetilde{\pi }}^{k}},\cdots,{{\widetilde{\pi }}^{n}} \right\}$, and the target policy ${\widetilde{\pi }}^{i}$ in each local environment $E^i$ is:
\begin{equation}\label{(target_pi)}
	\begin{split}
		{\widetilde{\pi }}^{i}=\underset{\pi^i}{\operatorname{argmax}} \sum_{t=0}^{T} E_{(s_{t}^i, a_{t}^i) \sim \tau_{\pi^i}}[\gamma r(s_{t}^i, a_{t}^i)+  \alpha^{i} \mathcal{H}(\pi^{i} (. \mid s_{t}^i))]
	\end{split}
\end{equation}
where ${\widetilde{\pi }}^{i}$ is the target policy, $\pi^i$ is a policy of agent $i$ in local environment $E^i$, $\gamma$ is the discount rate and $r$ is a reward from the environment, $s_t^i \in S$ and $a_t^i \in A$ denote state and action in local environment $E^i$ with timestamp $t$, and $S, A$ are global state space and global action space, respectively. $\tau_{\pi^i}$ is the distribution of trajectories generated from policy $\pi^i$, $\alpha^{i}$ is the temperature parameter to control the positivity of the policy exploration in the local environment and $\mathcal{H}(\cdot)$ indicates the entropy. And, in differentiated environments, the transfer probability varies among local environments, i.e. $P(s_{t+1}^i | s_t^i, a) \neq P(s_{t+1}^j | s_t^j, a), i \neq j$. \par

To obtain $\varPi$, the DFSAC algorithm uses an approach like the soft policy iteration\cite{haarnoja2018soft}, that alternately performs the two steps of policy evaluation and policy improvement to converge to the optimal value function and optimal policy. An example is that in agent k, the policy network takes the state $s_t^k$ of the local environment $E^k$ at moment $t$ as input to obtain the action $a^k_t$. To evaluate the impact of the policy on the local environment as well as on the global environment, the soft state value is defined as:
\begin{equation}\label{(V_function)}
	V\left(s_{t}^k\right):=\pi^k\left(s_{t}^k\right)^{T}\left[\chi \left(s_{t}^k\right)-\alpha^k \log \left(\pi^k\left(s_{t}^k\right)\right)\right]
\end{equation}
among them, $\chi$ denote the trend networks. And, we use the following loss function to update trend networks:
\begin{equation}\label{(trend_network_loss)}
	\begin{split}
		J_{\chi ^k}(\theta)=\mathbb{E}_{\left(s_{t}^k, a_{t}^k\right) \sim \mathcal{D}}\left[ \frac{1}{2}\Big(\chi_{\theta}^k\left(s_{t}^k, a_{t}^k\right)-\Big(r\left(s_{t}^k, a_{t}^k\right) \right.\Big.\Big. \\ \Big.\Big.\Big. +\gamma \mathbb{E}_{s_{t+1}^k \sim p}\left[V_{\bar{\theta}}\left(s_{t+1}^k\right)\Big]\Big)\Big)^{2}\right]
	\end{split}
\end{equation}
where $\mathcal{D}$ denotes a trajectory stored in replay memory and $V_{\bar{\theta}}$ denotes the use of global trend networks to calculate the soft
state value. This can be seen as a collaborative learning process using the global information shared by the collaborators. And the global trend networks are obtained by a soft update of each agent and then aggregated:
\begin{equation}\label{(soft_update)}
	\bar{\chi} \leftarrow \varepsilon \chi^k + (1-\varepsilon)\bar{\chi} , k \in \{1,2,\ldots,N\}
\end{equation}
where $\bar{\chi}$ is global trend networks, $\chi^k$ is local trend networks in agent $k$, $\varepsilon$ is the aggregation factor. Then, the updated trend networks are used to guide the policy improvement:
\begin{equation}\label{(pi_loss)}
	J_{\pi^k}(\phi)=E_{s_{t}^k \sim D}[\pi^k(s_{t}^k)^{T}[\alpha^k \log(\pi_{\phi}^k(s_{t}))-\chi_{\theta}^k(s_{t}^k)]]
\end{equation}
And $\alpha$ is the temperature parameter, but the appropriate value of $\alpha$ is different at different stages of training. Therefore, the selection of the {$\alpha$} is formulated as a constrained optimization problem\cite{haarnoja2018softaa} that maximizes the expected return while keeping the entropy of the policy greater than a threshold as follows:
\begin{equation}\label{(20)}
	\max _{\pi_{0}, \ldots, \pi_{T}} \mathbb{E}\left[\sum_{t=0}^{T} r\left(s_{t}, a_{t}\right)\right] \text { s.t. } \forall t, \mathcal{H}\left(\pi_{t}\right) \geq \mathcal{H}_{0}
\end{equation}
we also dynamically train with {$\alpha$} as a parameter of model, the loss function of {$\alpha^k$} in agent $k$ is\
\begin{equation}\label{(21)}
	J(\alpha^k)=\pi_{t}^k\left(s_{t}^k\right)^{T}\left[-\alpha^k\left(\log \left(\pi_{t}^k\left(s_{t}^k\right)\right)+\bar{H}\right)\right]
\end{equation}
\par
Through the continuous iteration of the above two steps, agent $K$ is guided by global trend networks to obtain the target policy ${\widetilde{\pi }}^{k}$ applicable to the local environment $E^k$ by sharing knowledge with other agents eventually.\par

The complete DFSAC algorithm we proposed is shown in Algorithm 1.
Firstly, the algorithm initializes the local network parameters of local networks (two local trend networks and one policy network).
Then initialize the global network (two global trend networks), and equalize the global trend networks and local trend networks' parameters.
The agents first initialize an empty replay memory and store a backup of the global trend networks.
And the agent will collect information about the local environment as the input of the policy network and output an action.
Then, the agent gets the reward value and the next state from the local environment calculates the cumulative discount function at each moment and stores the transition in the replay memory.
When the number of transitions in the replay memory reaches the number set in advance, the algorithm starts training.
It updates the local network parameters. Finally, the algorithm updates the temperature parameters.
After running preset training iterations, the agents will upload their local trend network parameters to the federated center node.
After receiving the local trend network parameters of agents, the federated center node integrates these parameters and updates the global trend network parameters.
Afterward, the federated center node distributes the global trend network parameters to each agent, and agents update their backup with the latest global trend networks.
Through this integrated and distributed processing method, the local network in the agents can reflect the real-time situation of the global environment.

\begin{algorithm}[htbp]
	\caption{DFSAC algorithm}
	\begin{algorithmic}[1]
		\State Initialize $\mathrm{\chi}_{\theta_{1}}^n: S \rightarrow \mathbb{R}^{|A|}, \mathrm{\chi}_{\theta_{2}}^n: S \rightarrow \mathbb{R}^{|A|}, \pi_{\phi}^n: S \rightarrow[0,1]^{|A|}$ for $n\in \left\{ 1,2,\ldots ,N \right\}$ 
		\Comment{Initialize local network parameters}
		\State Initialize $\overline{\mathrm{\chi}}_{\theta_{1}}: S \rightarrow \mathbb{R}^{|A|}, \overline{\mathrm{\chi}}_{\theta_{2}}: S \rightarrow \mathbb{R}^{|A|}$
		\Comment{Initialize global trend networks parameters at the federated center node}
		\State $\theta_{1}^n \leftarrow \bar{\theta}_{1}, \theta_{2}^n \leftarrow \bar{\theta}_{2}$ for $n\in \left\{ 1,2,\ldots ,N \right\}$
		\Comment{Equalize global trend networks and local trend network parameters}
		\State $D^n \leftarrow \varnothing$ for $n\in \left\{ 1,2,\ldots ,N \right\}$
		\Comment{Initialize an empty replay memory}
		\While{running}
		\For {each agent $n$}
		\State Get state ${{s}_{t}}$ from the environment $E^n$
		\State ${{a}_{t}}\sim\pi \left( {{a}_{t}}|{{s}_{t}} \right)$
		\Comment{Sample action from the agent $n$}
		\State $s_{t+1} \sim p\left(s_{t+1} \mid s_{t}, a_{t}\right)$
		\Comment{Sample transition from the environment $E^n$}
		\State $D^n\leftarrow D^n\cup \left\{ \left( {{s}_{t}},{{a}_{t}},r\left( {{s}_{t}},{{a}_{t}} \right),{{s}_{t+1}} \right) \right\}$
		\Comment{Store the transition in replay memory}
		\State ${{\theta }_{i}^n}\leftarrow {{\theta }_{i}^n}-{{\lambda }_{\chi}}{{\hat{\nabla }}_{{{\theta }_{i}^n}}}{{J}_{\chi}}\left( {{\theta }_{i}^n} \right)$ for $i\in \left\{ 1,2 \right\}$ and $n\in \left\{ 1,2,\ldots ,N \right\}$
		\Comment{Update local trend networks use Eq~\eqref{(trend_network_loss)}}
		\State $\phi^n \leftarrow \phi^n -{{\lambda }_{\pi }}{{\hat{\nabla }}_{\phi ^n}}{{J}_{\pi }}\left( \phi^n  \right)$ for $n\in \left\{ 1,2,\ldots ,N \right\}$
		\Comment{Update policy networks use Eq~\eqref{(pi_loss)}}
		\State $\alpha^n \leftarrow \alpha^n -\lambda {{\hat{\nabla }}_{\alpha^n }}J\left( \alpha^n  \right)$
		\Comment{Update temperature use Eq~\eqref{(21)}}
		\State When running k iterations upload ${{\theta }_{1}^n},{{\theta }_{2}^n}$ to federated center node
		\EndFor
		\If{in federated center node}
		\State ${{\bar{\chi}}_{i}}\leftarrow \varepsilon \chi_{i}^{n}+(1-\varepsilon ){{\bar{\chi}}_{i}}$ for $i\in \left\{ 1,2 \right\}$ and $n\in \left\{ 1,2,\ldots ,N \right\}$
		\Comment{Aggregated global trend network use Eq.~\eqref{(soft_update)}}
		\EndIf
		\EndWhile
	\end{algorithmic}
\end{algorithm}


\begin{figure*}[htbp]
	\centering
	\includegraphics[width=0.8\linewidth]{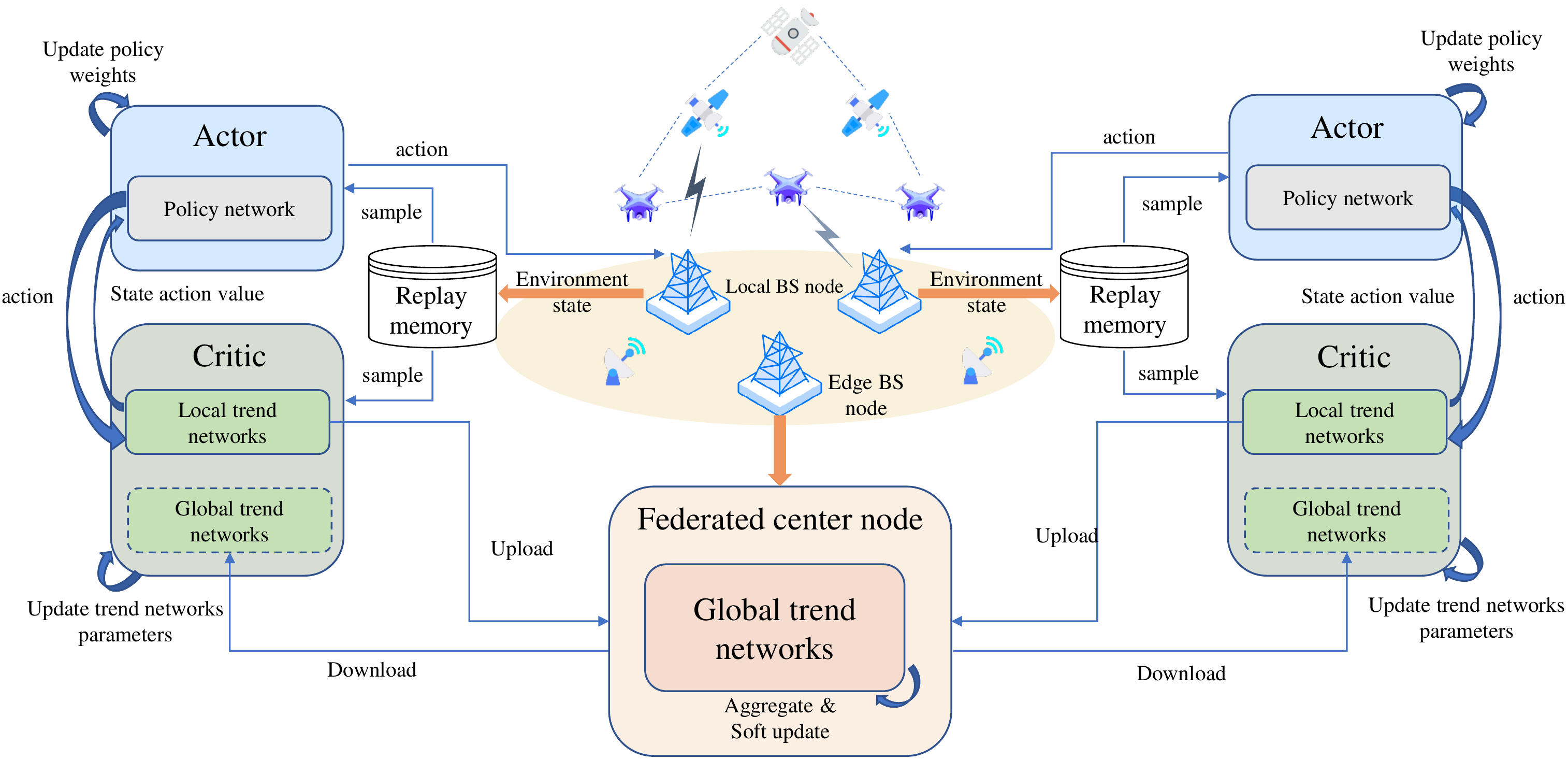}
	\caption{The frame structure of DFSAC-based traffic offloading method in SAGIN.}
	\label{fig:Model}
\end{figure*}

\section{Empirical Study}
\subsection{Settings}
In this section, we describe the simulation environment and evaluate our DFSAC-based traffic offloading method. The simulation environment is a dynamic four-layer SAGIN environment, which divides each device into three types: the source node device type, the relay node device type, and the destination node device type.\par
The source node device is used to generate data packets, which are transmitted to the destination node device for reception. The source node device and the destination node device are modeled as user equipment (UE). The source node device generates data using a normally distributed data generation rate. 
The relay node device transfers data packets based on the Open Shortest Path First (OSPF) traffic routing protocol or the proposed traffic offloading method. 
The relay nodes are modeled as BS, UAV, LEO, and GEO. Except for BSs are static, other devices have their mobile models. 
The source node and destination node devices use a random-waypoint mobility model \cite{soltani2018modeling}, which can simply randomize the device's location. \par

In our simulation, the UEs use the arbitrary movement model to move within a $100km^2$ two-dimensional plane area. For the mobility model of UAV, we set the UAV to fly in a certain direction at a fixed speed in a predefined period. 
After a while, the UAV's moving direction will shift arbitrarily and fly at the same moving speed. We consider that LEO periodically covers the considered area as it moves around the earth. For GEO, due to its high coverage, we consider it to cover the target area all the time and have a stable connection (i.e., the air condition change is not considered). For this area, $(\varphi, \omega_{0}, \eta, \gamma, k_{0})$ is set to (3.04, -3.61, -23.29, 4.14, 20.7) \cite{zhou2019delay}. For the rain attenuation of the satellite channel model, we set the $F_{rain}$ to 6dB \cite{zhou2019delay}. The remaining simulation parameters are shown in Table I.\par

\begin{table}[htbp]
	\centering
	\caption{Simulation Parameters}
	\begin{tabular}{ll}
		\toprule
		Parameters                  & Values                 \\
		\midrule
		Number of BS nodes          & 8                      \\
		Number of UAV nodes         & 6                      \\
		Number of LEO nodes         & 2                      \\
		Number of GEO nodes         & 1                      \\
		Packet generation rate      & $N(1e6,\sigma^2)$ Mbps \\
		BS - UAV bandwidth          & 20MHz                  \\
		BS - LEO bandwidth          & 37.5MHz                \\
		BS - GEO bandwidth          & 25MHz                  \\
		UAV - LEO bandwidth         & 10MHz                  \\
		UAV - GEO bandwidth         & 5MHz                   \\
		Gamma                       & 0.99                   \\
		{$\varepsilon$}             & 1e-2                   \\
		Learning rate               & 5e-4                   \\
		Learning rate of {$\alpha$} & 1e-3                   \\
		Target entropy              & -4.0                   \\
		\bottomrule
	\end{tabular}%
	\label{tab:addlabel}%
\end{table}%

The frame structure of the DFSAC-based traffic offloading method is shown in Fig.~\ref{fig:Model}. 
The local base station nodes are training nodes for federated learning to collect the network environment state in replay memory for training. 
The neural network in each training node can be divided into Actor-network and Critic-network. 
The Actor-network is responsible for outputting the action of traffic offloading based on the policy network. 
The Critic network is responsible for judging the actions and the environmental trends, which are used to optimize and improve the performance of the policy network. 
Meanwhile, the edge base station node is used as a federated center node to receive local trend networks uploaded by the local base station nodes, aggregate the parameters, and then distribute them to the local base station nodes to update their network parameters to the latest.\par

\subsection{Performance Evaluation}
To evaluate our proposed traffic offloading method, we conducted experiments comparing it with existing approaches. Specifically, we run the traditional traffic offloading method used in \cite{kang2014mobile,yu2017mobile,ding2015vision,korhonen2013toward} in the same environment. The traditional traffic offloading method is a greedy offloading method based on the shortest path.  In addition, we simulated the distributed DDQN-based traffic offloading method in \cite{tang2021deep} and the traditional FRL method for experimental comparison.\par

\begin{figure}[htbp]
\centering
\includegraphics[width=2.5in]{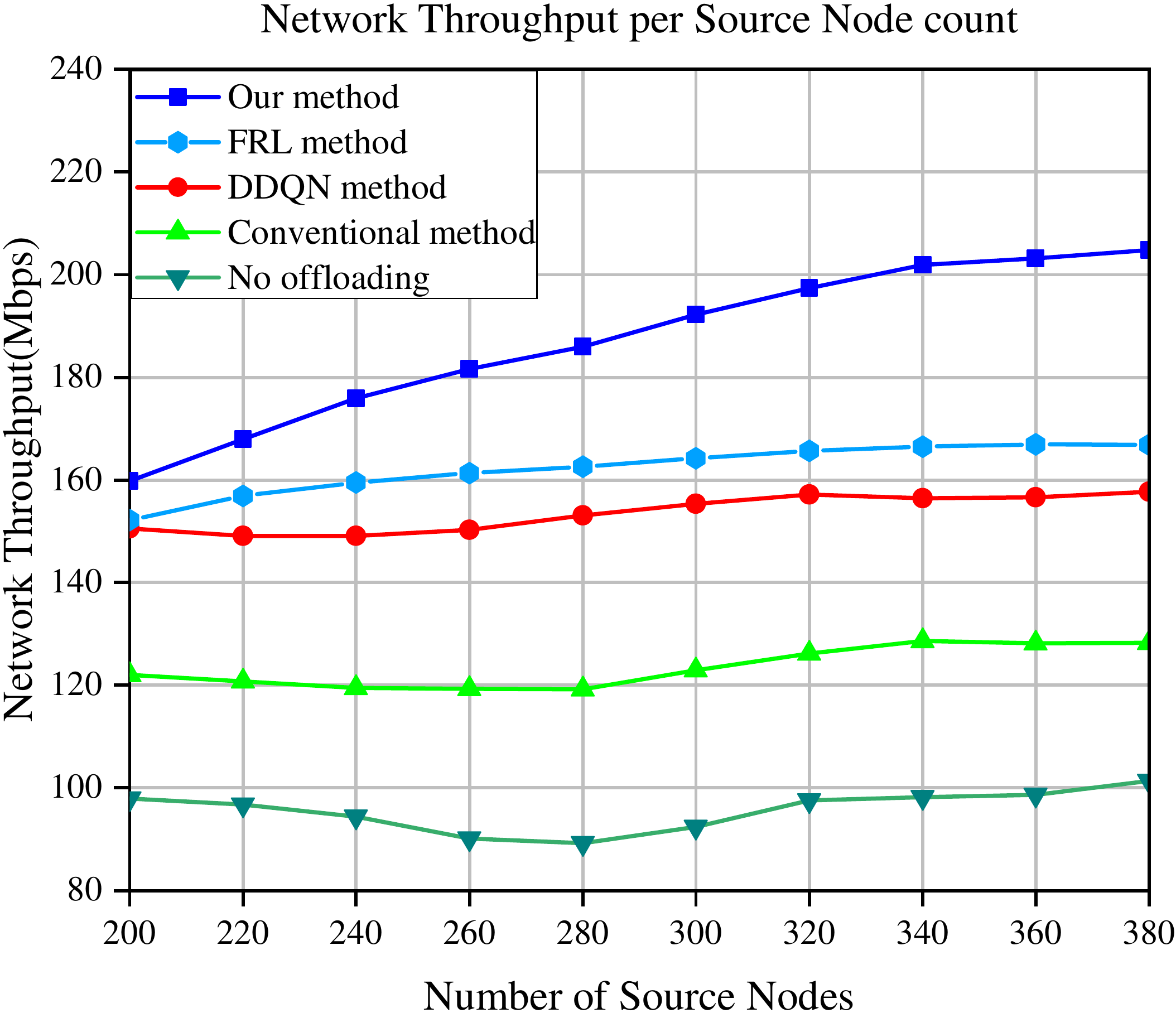}
\caption{ Throughput per increasing source node count. }
\label{fig:throughput}
\end{figure}

\begin{figure}[htbp]
\centering
\includegraphics[width=2.5in]{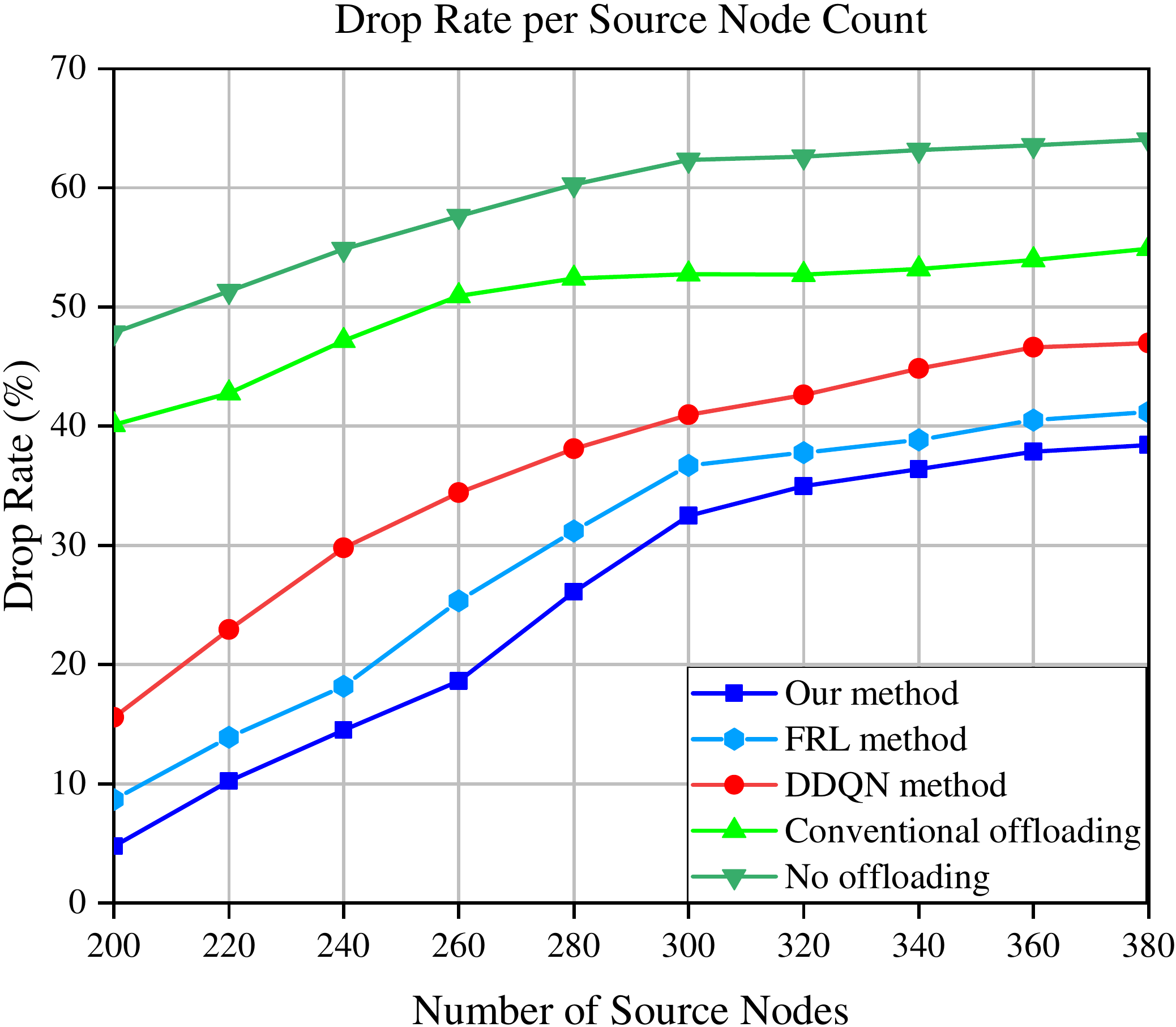}
\caption{ Drop rate per increasing source node count. }
\label{fig:droprate}
\end{figure}

Fig.~\ref{fig:throughput} and Fig.~\ref{fig:droprate} present a comprehensive analysis of throughput and packet loss rate variations as the number of source nodes incrementally increases from 200 to 380, involving the evaluation of four distinct methods. Fig.~\ref{fig:throughput} clearly demonstrates the substantial advantages of our proposed method. As the number of source nodes increases, our method exhibits a consistent improvement in throughput, contrasting with the unaltered performance of the other methodologies. This compelling observation underscores our method's remarkable ability to enhance network throughput and maximize its capacity compared to its counterparts. Fig.~\ref{fig:droprate} shows the dynamics of packet loss rates as the source node count escalates. It becomes evident that, with the increasing number of source nodes, the packet loss rates of all methods tend to rise, a common trend attributed to network congestion. However, our proposed method emerges as the standout performer, consistently maintaining a lower packet loss rate in comparison to the other three methods. This noteworthy outcome underscores the efficacy of our approach in mitigating packet loss and ensuring reliable data transmission in dynamic and challenging network environments.

\begin{figure}[htbp]
\centering
\includegraphics[width=2.5in]{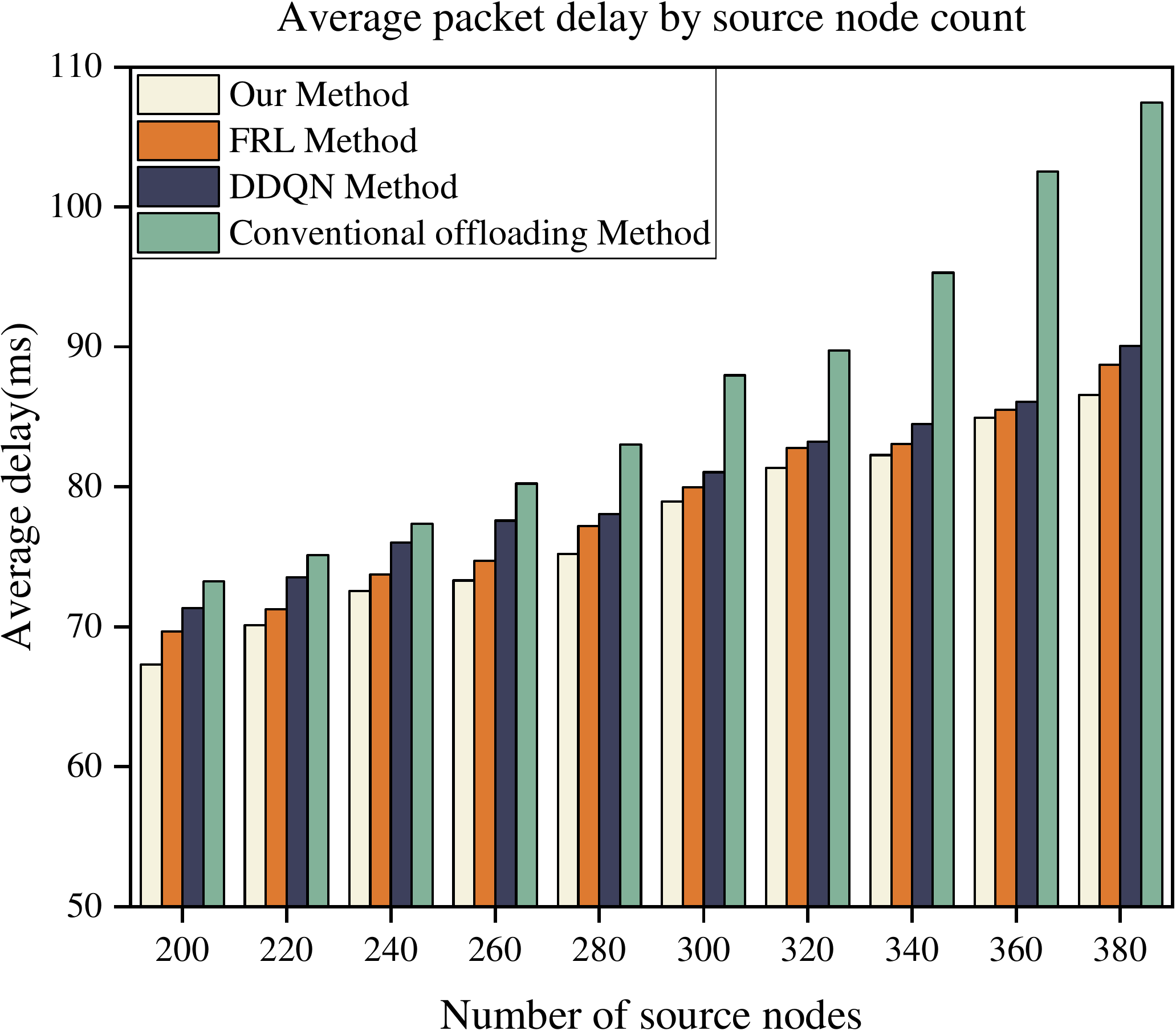}
\caption{Packet delay per increasing source node count. }
\label{fig:delay}
\end{figure}

Next, we focus on evaluating packet delay performance, as depicted in Fig.~\ref{fig:delay}, when these methods are deployed while varying the number of source nodes. A clear trend emerges from the figure: when the offloading method is not utilized, there is a noticeable and consistent increase in average packet delay as the number of source nodes grows. The strategy proposed in this study significantly outperforms the methods based on DDQN and FRL in scenarios with source node counts ranging from 200 to 380. This empirical evidence highlights the remarkable capability of our method to maintain the lowest delay, even amid increasing network loads and a significant rise in the number of nodes. Notably, such performance demonstrates the robust stability of our approach in maintaining low latency within dynamic heterogeneous network environments.

\begin{figure}[htbp]
\centering
\includegraphics[width=2.5in]{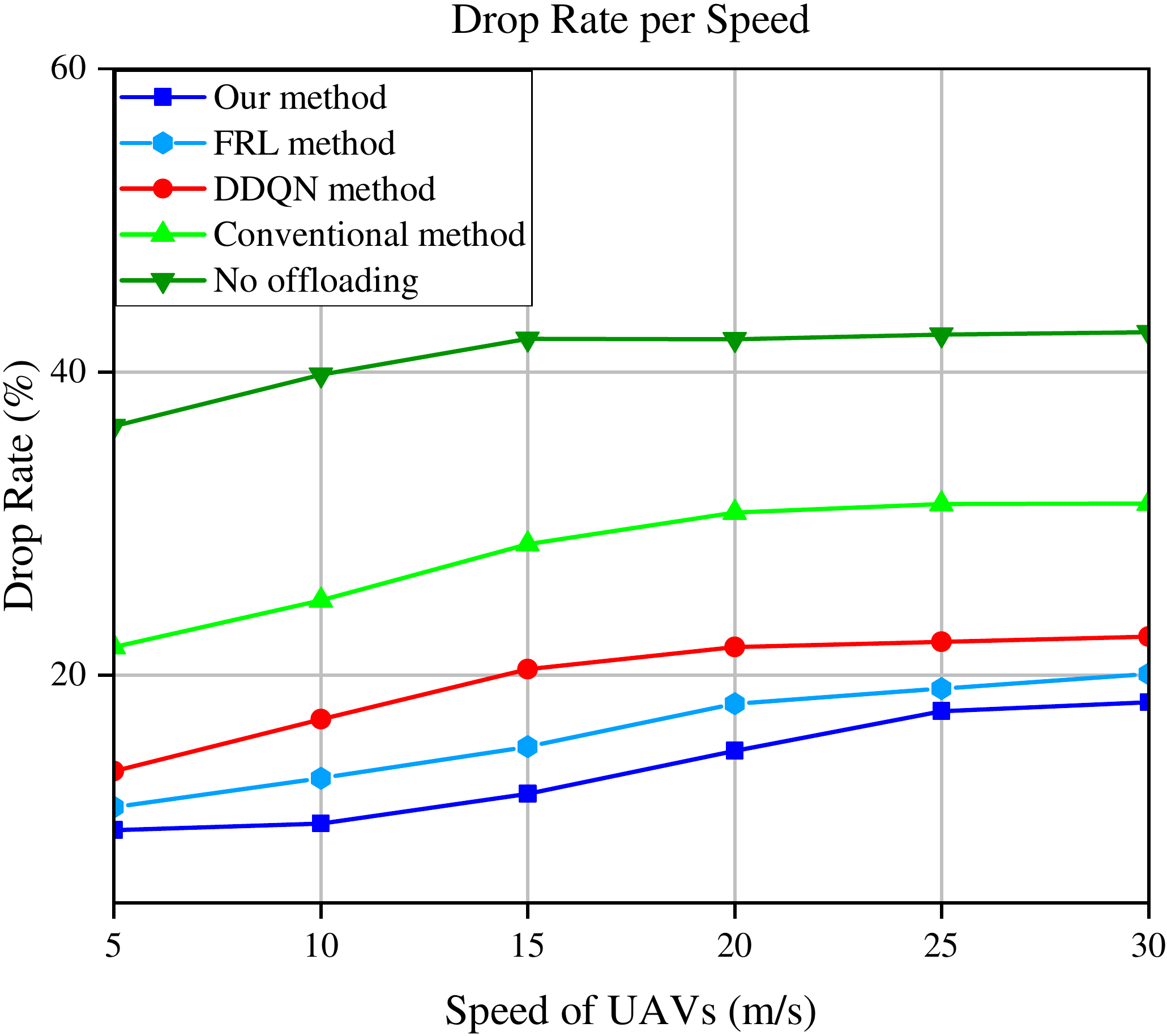}
\caption{Packet drop rate with increasing moving speed of UAVs. }
\label{fig:droprateSpeed}
\end{figure}

\begin{figure}[htbp]
\centering
\includegraphics[width=2.5in]{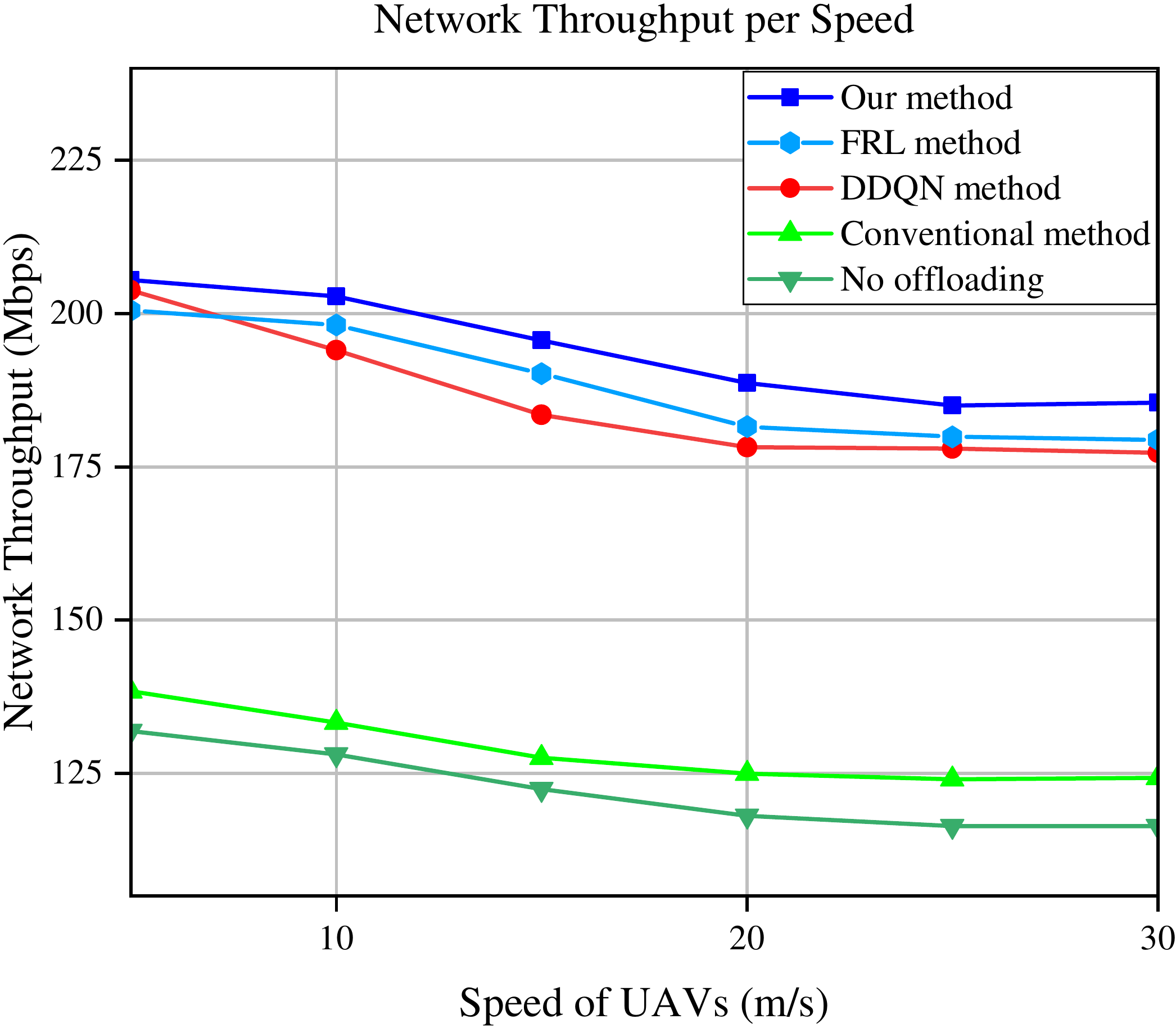}
\caption{Throughput with increasing moving speed of UAVs. }
\label{fig:throughputSpeed}
\end{figure}

Further, to assess the robustness of our algorithm in the high-mobility setting of SAGIN, we conducted simulations to evaluate packet drop rates and throughput. The movement characteristics of the satellites are predetermined within the environment. Therefore, we systematically varied the speeds of the UAVs, ranging from 5m/s to 30m/s, to assess the performance of each method. Fig.~\ref{fig:droprateSpeed} and Fig.~\ref{fig:throughputSpeed} show the packet drop rate increases, and throughput decreases with UAV moving speed increase. Nevertheless, our proposed method consistently outperforms the alternatives even as the UAV speed increases. This trend aligns with the earlier experimental results, affirming the overall superiority of our proposed traffic offloading method in the context of SAGIN, particularly in high-mobility scenarios.\par

\subsection{Validity of DFRL Framework}
The distributed DDQN-based traffic offloading approach \cite{tang2021deep} is at the forefront of achieving optimal performance in highly dynamic SAGIN environments. Building upon this foundation, we undertake a comprehensive evaluation to validate the effectiveness of our proposed DFRL framework thoroughly. We meticulously compare and contrast the performance of three distinct approaches: the distributed DDQN-based traffic offloading approach \cite{tang2021deep}, the FL-based DDQN approach, and the DFRL-based DDQN approach. 

\begin{figure}[htbp]
\centering
\includegraphics[width=2.5in]{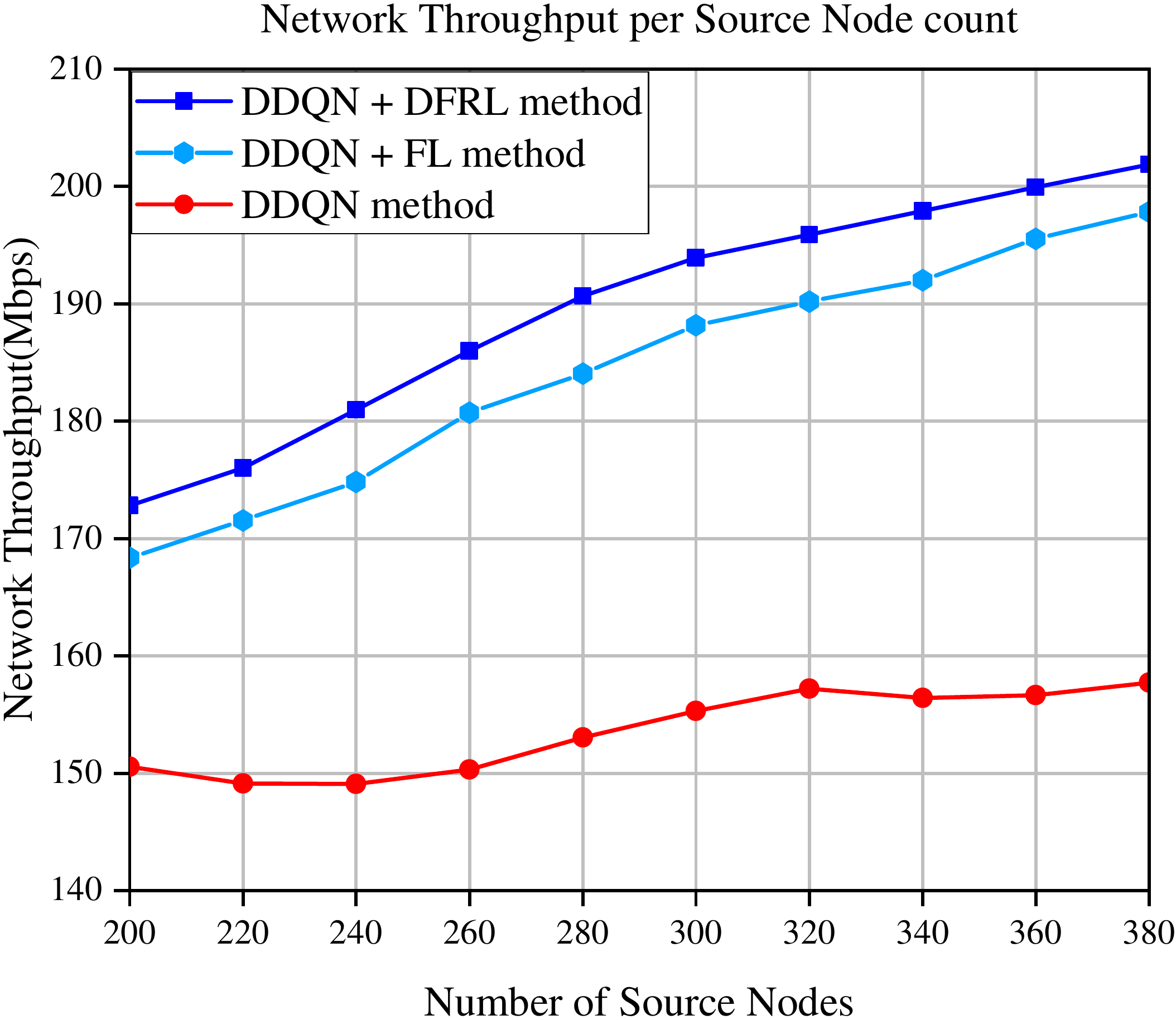}
\caption{ Throughput per increasing source node count. }
\label{fig:DDQNthroughput}
\end{figure}

\begin{figure}[htbp]
\centering
\includegraphics[width=2.5in]{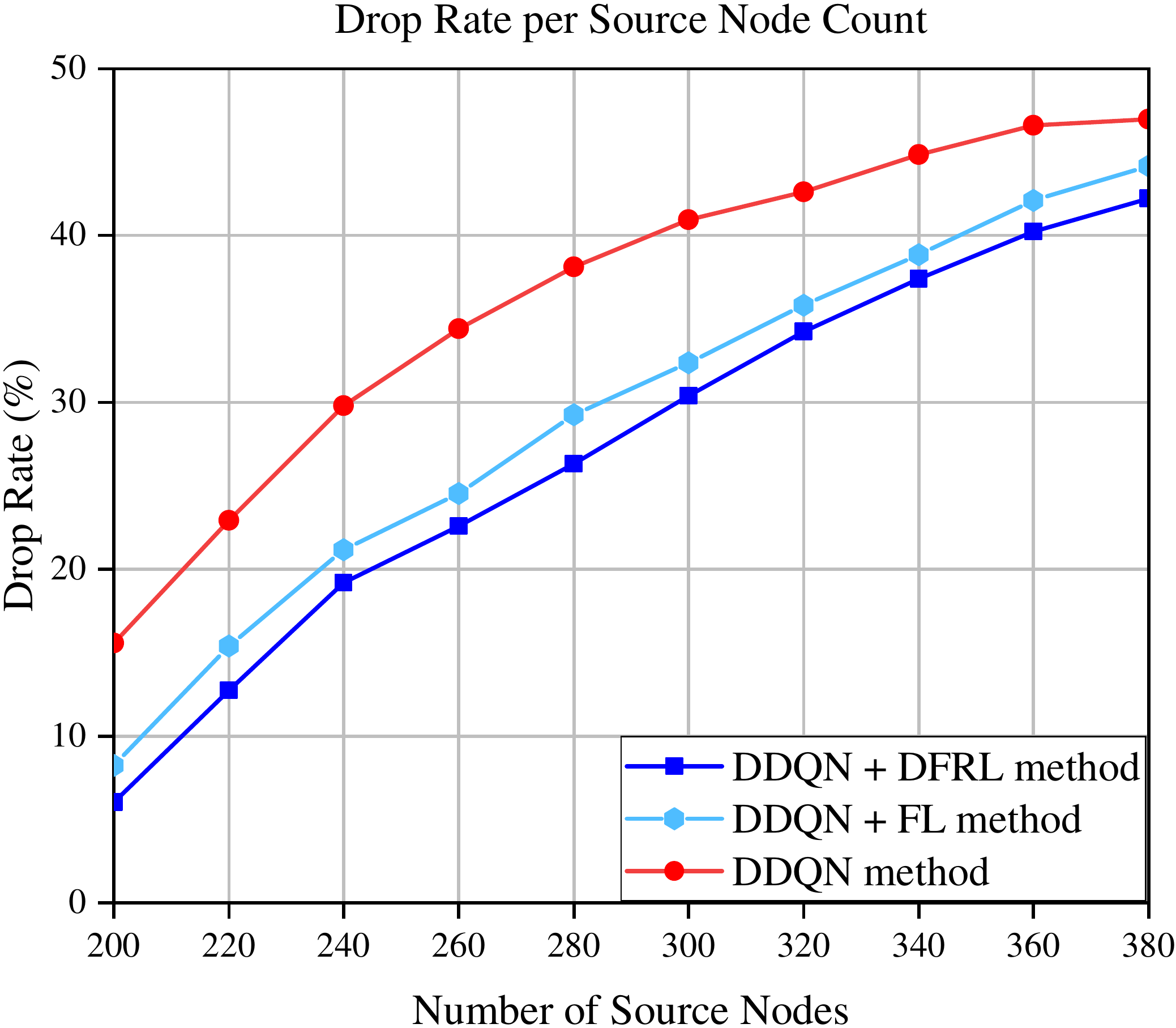}
\caption{ Drop rate per increasing source node count. }
\label{fig:DDQNdroprate}
\end{figure}

\begin{figure}[htbp]
\centering
\includegraphics[width=2.5in]{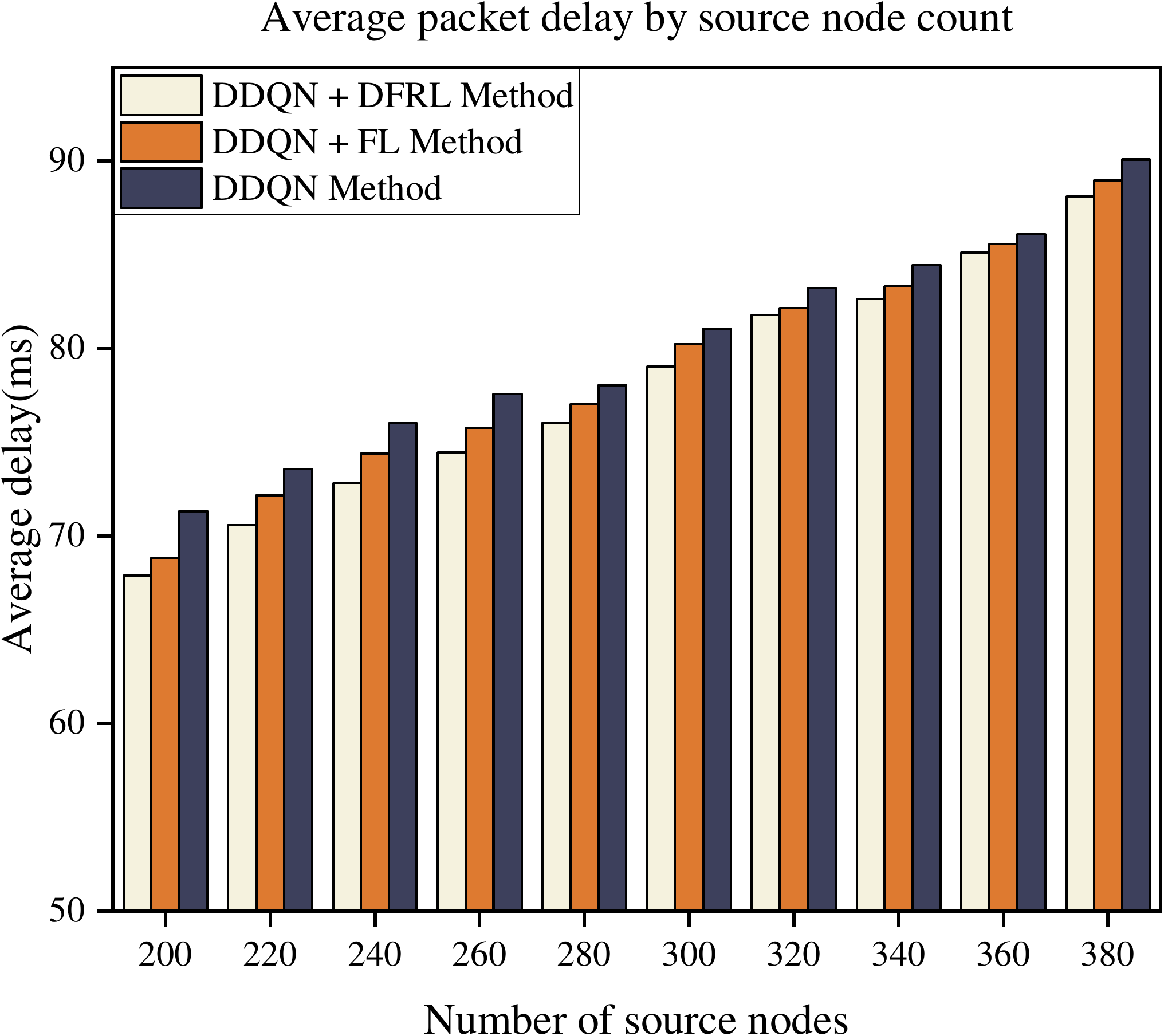}
\caption{Packet delay per increasing source node count. }
\label{fig:ddqndelay}
\end{figure}

Fig.~\ref{fig:DDQNthroughput}, Fig.~\ref{fig:DDQNdroprate} and Fig.~\ref{fig:ddqndelay} show throughput, packet loss rate, and packet delay fluctuation as the number of source nodes increases gradually from 200 to 380. The graphical representation clearly and decisively showcases the superior performance of the FL-based DDQN approach compared to the approach detailed in \cite{tang2021deep}. This advantage can be attributed to the heightened interaction among intelligent agents, which empowers them to make more effective decisions by harnessing global state information, resulting in consistently superior outcomes. Furthermore, as the number of source nodes continuously increases, leading to a corresponding rise in network load, the DFRL-based DDQN approach consistently outperforms the other two algorithms, exhibiting exceptional performance across three key performance indicators. This is attributed to our proposed DFRL framework, where agents develop an understanding of the uniqueness of each region based on information collected from multiple areas, thereby formulating more suitable local strategies for their respective regions. This serves as a demonstration of the robustness and effectiveness of our proposed DFRL framework, providing a viable solution to address the intricate challenges posed by the dynamic and heterogeneous SAGIN environment.


\subsection{Application Cases}
We use CartPole from OpenAI Gym \cite{brockman2016openai} as the experimental environment.
CartPole is a cart-pole game with a cart and a pole erected.
The algorithm must control the cart to move left or right to keep the pole upright while satisfying the constraints.
Whenever an operation is performed without tilting the rod beyond the limit, the environment rewards the agent with a value of 1, it otherwise 0. Our goal is to verify the algorithm's performance in several differentiated environments. Therefore, the transfer probability of the environment is varied by changing the length of the pole to obtain multiple differentiated environments and setting up agents in each of the differentiated environments.\par

For the DFSAC algorithm, we use the output of the policy network to control the agents.
An additional node is set as a federated center node for model aggregation and distribution.
As described in the previous section, agents in different environments share knowledge by sharing global trend networks.
Considering the communication overhead between agents, each agent stores a backup copy of global trend networks locally to guide the local data training in the actual algorithm implementation.
After a certain number of training sessions, i.e., soft update, the local trend networks are uploaded to the federal center node for aggregation, and the latest global trend networks are downloaded to update the local backup.
This approach can reduce the communication overhead incurred during training.\par

The characteristic between the DFSAC algorithm and the traditional FRL algorithm is that the differentiation between different environments is considered in the learning process.
Therefore, we use the SAC algorithm trained by traditional federated learning as our baseline, i.e. FedAVG~\cite{mcmahan2017communication}, the agents in each environment share the model, including both value parameters and policy model parameters in the SAC algorithm, and the models are aggregated and distributed periodically using averaging model parameters.
In addition, we implement a centralized RL algorithm to verify the performance of a non-federated learning algorithm in this scenario. The centralized RL algorithm uses a SAC agent to collect information directly in each environment and then train.\par

\begin{figure}[htbp]
\centering
\includegraphics[width=2.5in]{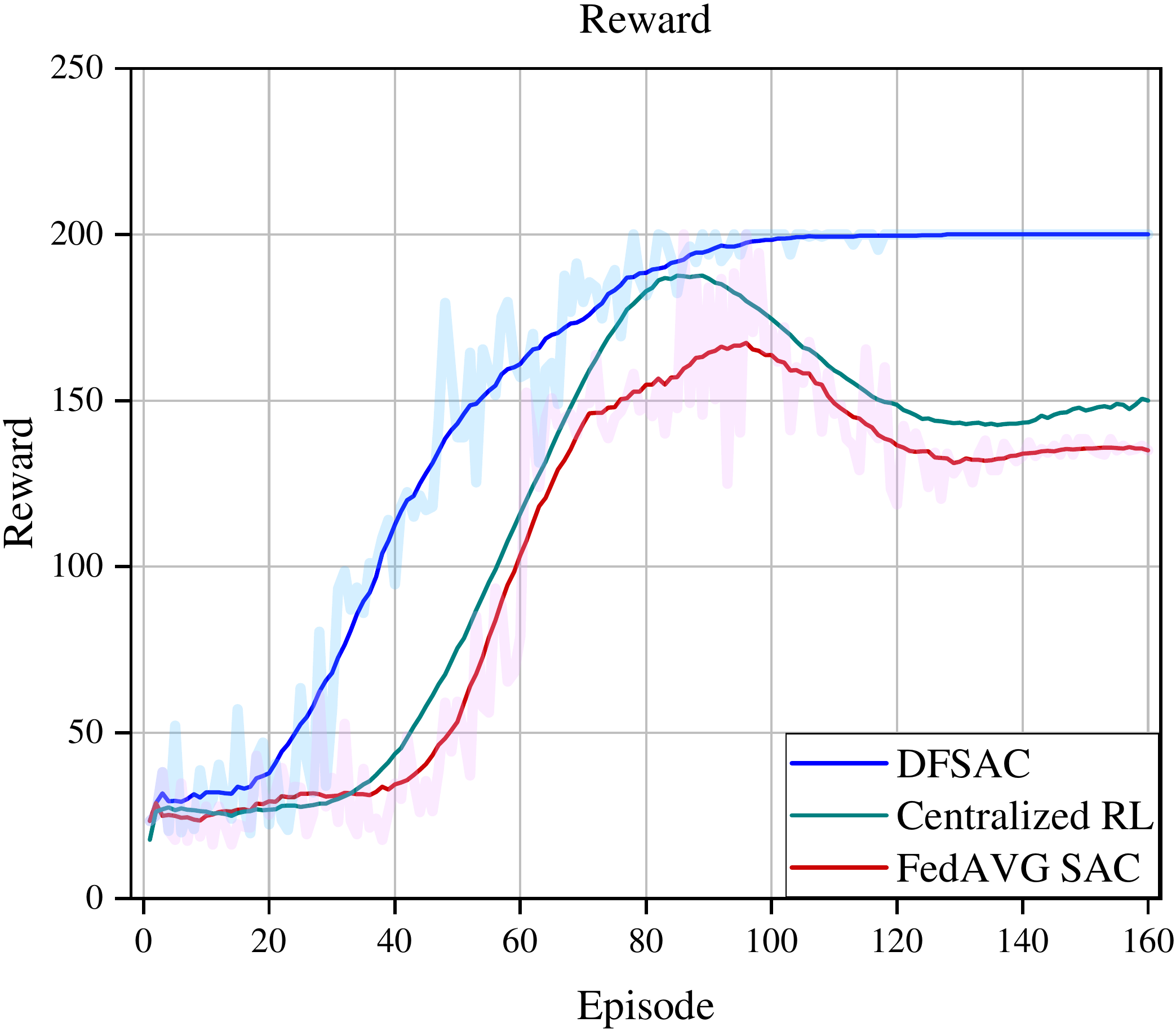}
\caption{The rewards achieved by different algorithms. }
\label{fig:reward_cp}
\end{figure}

\begin{figure}[htbp]
\centering
\includegraphics[width=2.5in]{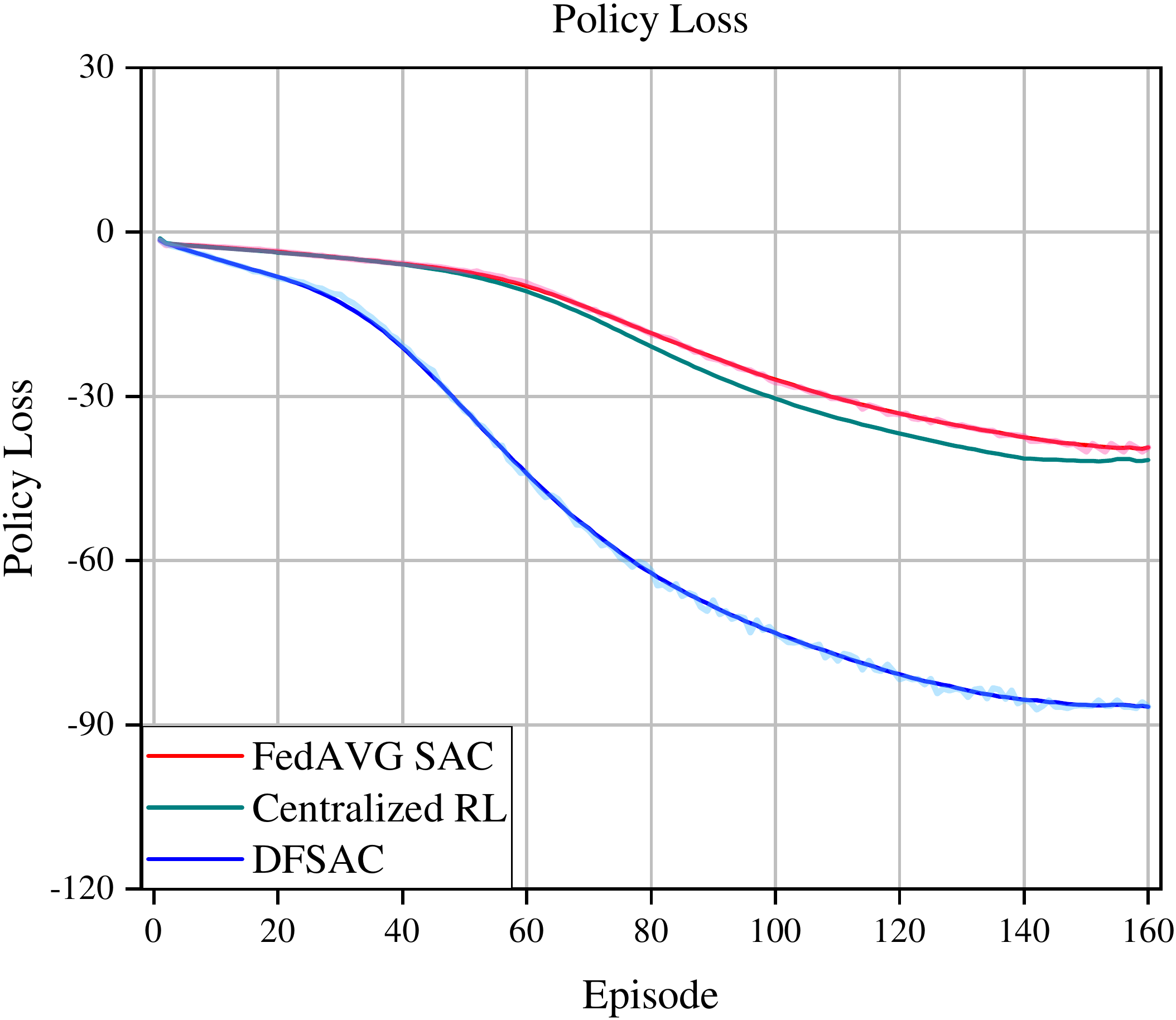}
\caption{The policy network loss evolution. }
\label{fig:loss_cp}
\end{figure}

The reward obtained by the different algorithms in each episode is shown in Fig.~\ref{fig:reward_cp}, which results from averaging the reward from several differentiated environments.
As can be seen, the DFSAC algorithm steadily increases and eventually converges as the training progresses.
In contrast, the reward of other algorithms fluctuates significantly during the training process and is lower than that of the DFSAC algorithm.
Fig.~\ref{fig:loss_cp} shows the change in the average loss value of each agent's policy network during the training process.
It can be seen from the figure that the DFSAC algorithm has higher learning efficiency than others.
This indicates that the DFSAC algorithm can obtain policy models more suitable for agents in differentiated environments than the traditional FRL algorithm.
The global policy model obtained by directly aggregating the traditional FRL algorithm is difficult to adapt to this differentiated environment.
Further, the DFSAC algorithm can avoid the additional communication overhead incurred by the centralized RL algorithm when transmitting environmental information.\par


\section{Conclusion}
The dynamic and heterogeneous nature of the environment gives rise to distinct regions, creating a potential challenge for RL due to the risk of getting trapped in local strategies. This study introduces a novel concept called DFRL tailored for networks operating in dynamic and heterogeneous environments. DFRL isolates the local policy model from global integration and employs a trend model to discern regional variations. In contrast to conventional FRL, our proposal strongly emphasizes adjusting biases in differentiated regions and ensuring the independence of local policy models. Our approach demonstrates superior performance, particularly in scenarios characterized by heterogeneous network environments.

\bibliographystyle{IEEEtran}
\bibliography{mian}
\begin{IEEEbiography}
[{\includegraphics[width=1in,height=1.25in,clip,keepaspectratio]{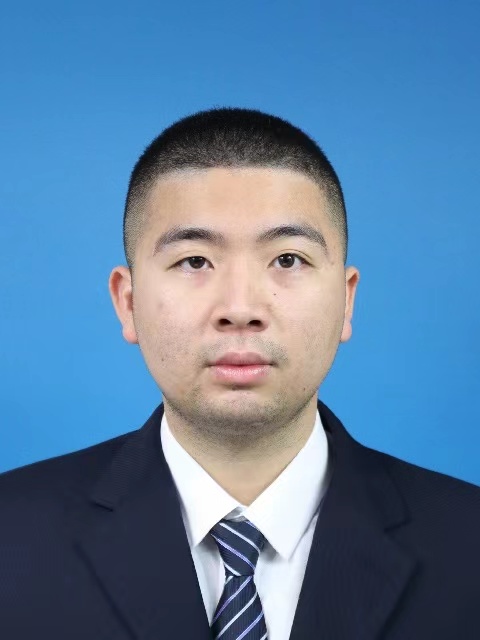}}]
\footnotesize \textbf{Yeguang Qin} received the M.E. degree from the Department of Software Engineering, Xinjiang University, in 2023. He is currently the Ph.D. Student with the Department of Computer Science and Technology, Central South University, advised by Prof. Ming Zhao. His research work focuses on wireless networks, network traffic control, and machine learning algorithm.
\end{IEEEbiography}

\begin{IEEEbiography}
[{\includegraphics[width=1in,height=1.25in,clip,keepaspectratio]{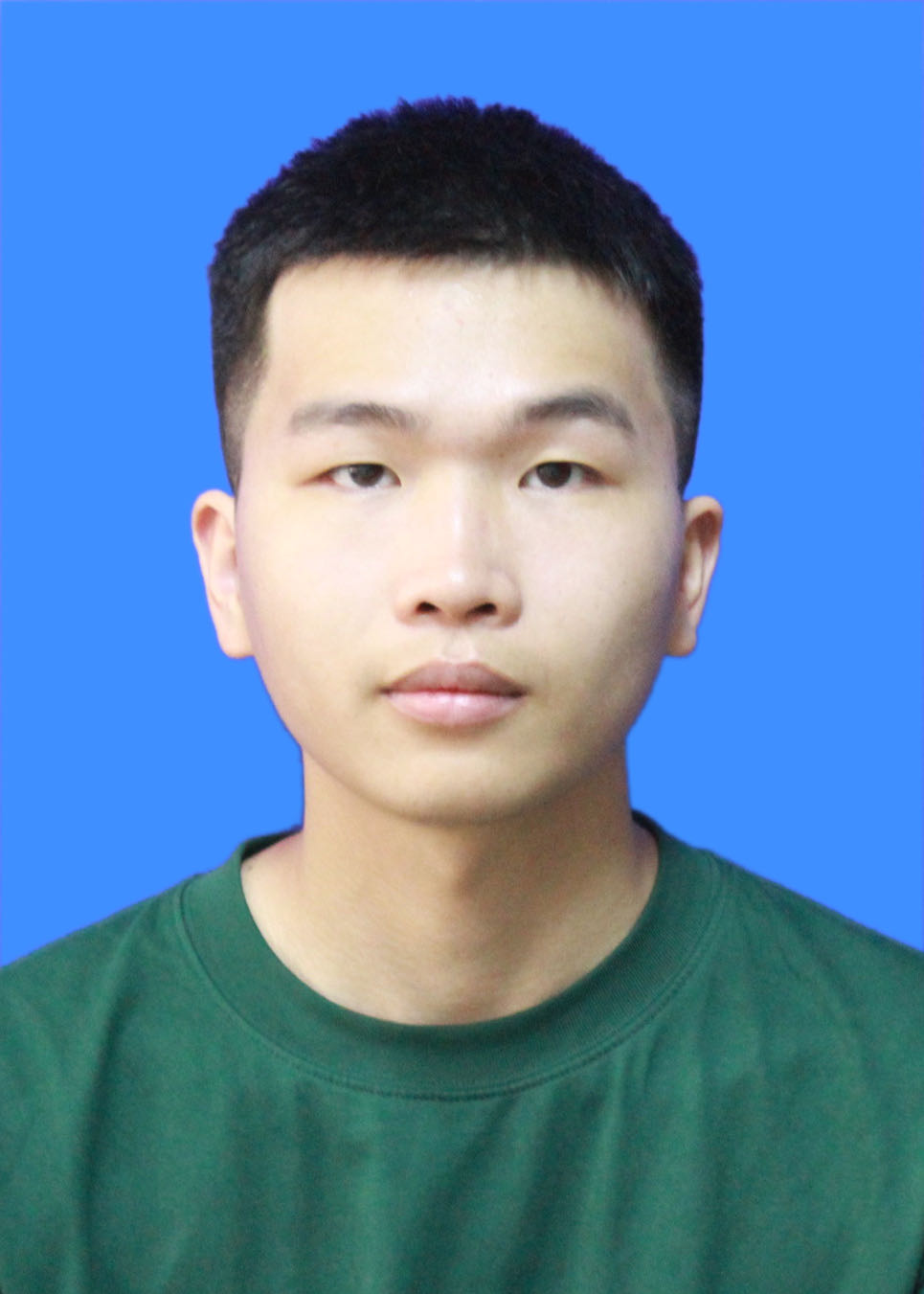}}]
\footnotesize \textbf{Yilin Yang} received the M.E. degree from the Department of Computer Science and Technology, Northeast Forestry University, in 2020 and the M.S. degree in software engineering from the School of Computer Science and Engineering, Central South University, Changsha, China, in 2023. His research work focuses on network traffic control and machine learning algorithm.
\end{IEEEbiography}

\begin{IEEEbiography}
[{\includegraphics[width=1in,height=1.25in,clip,keepaspectratio]{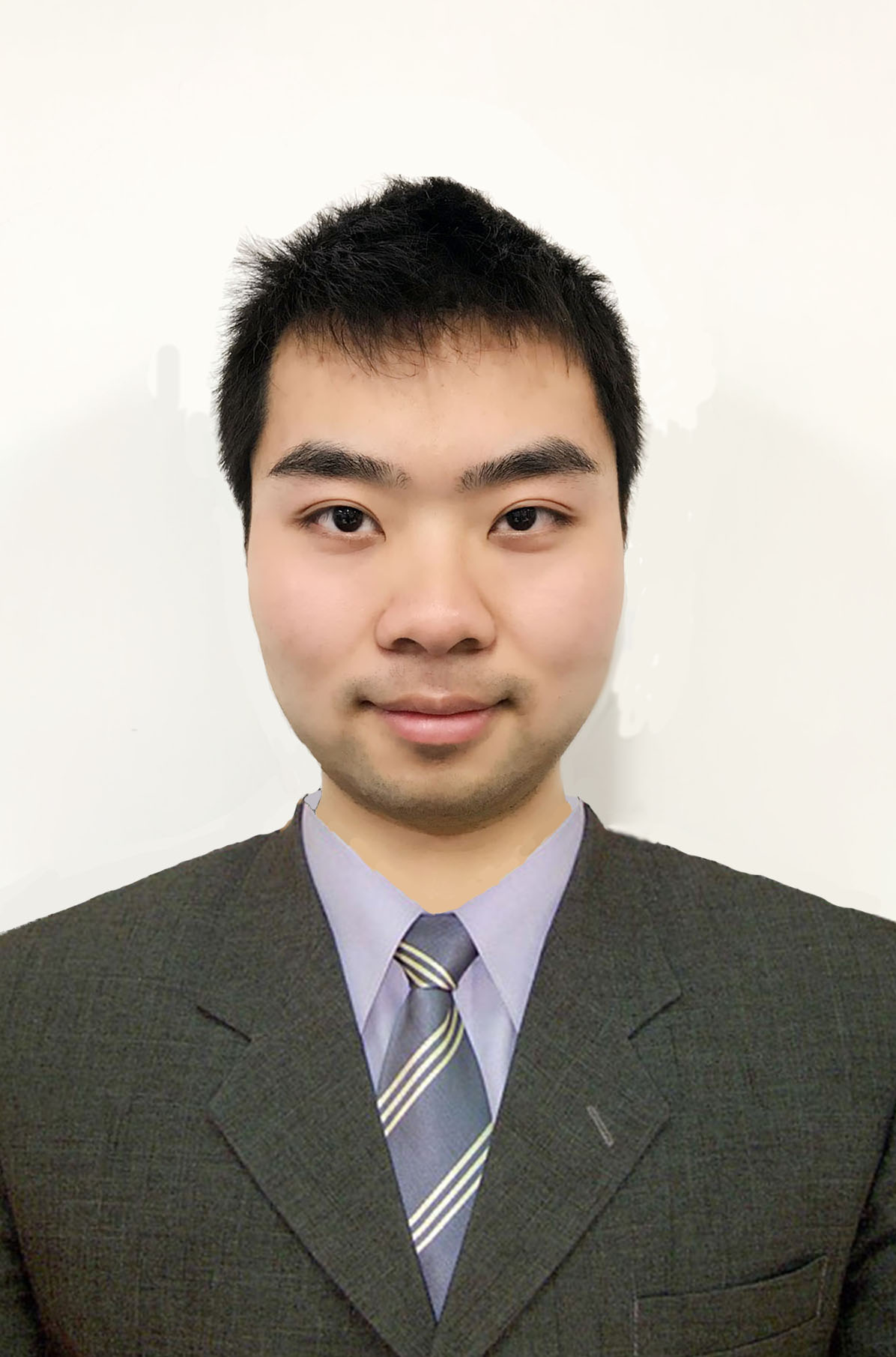}}]
\footnotesize \textbf{Fengxiao Tang}
received the B.E. degree in measurement and control technology and instrument from the Wuhan University of Technology, Wuhan, China, in 2012 and the M.S. degree in software engineering from the Central South University, Changsha, China, in 2015. He received the Ph.D. degrees from the Graduate School of Information Science, Tohoku University, Japan. Currently, He is an full professor in the School of Computer Science and Engineering of Central South University. He has been an Assistant Professor from 2019 to 2020 and an Associate Professor from 2020 to 2021 at the Graduate School of Information Sciences (GSIS) of Tohoku University. His research interests are unmanned aerial vehicles system, IoT security, game theory optimization, network traffic control and machine learning algorithm. He was a recipient of the prestigious Dean's and President's Awards from Tohoku University in 2019, and several best paper awards at conferences including IC-NIDC 2018, GLOBECOM 2017/2018. He was also a recipient of the prestigious Funai Research Award in 2020, IEEE ComSoc Asia-Pacific (AP) Outstanding Paper Award in 2020 and IEEE ComSoc AP Outstanding Young Researcher Award in 2021.
\end{IEEEbiography}

\begin{IEEEbiography}
[{\includegraphics[width=1in,height=1.25in,clip,keepaspectratio]{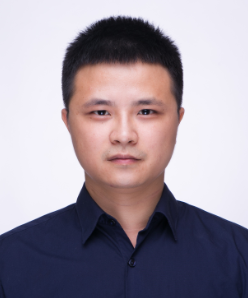}}]
\footnotesize \textbf{Xin Yao}
received the B.S. degree
in computer science from Xidian University, Xi’an,
China, in 2011, the M.S. degree in software engineering and the Ph.D. degree in computer science and
technology from Hunan University, Changsha, China,
in 2013 and 2018, respectively.
From 2015 to 2017, he worked as a Visiting Scholar
with Arizona State University. He is currently an
Assistant Professor with Central South University.
His research interests include security and privacy
issues in social network, Internet of Things, cloud
computing, and big data
\end{IEEEbiography}

\begin{IEEEbiography}
[{\includegraphics[width=1in,height=1.25in,clip,keepaspectratio]{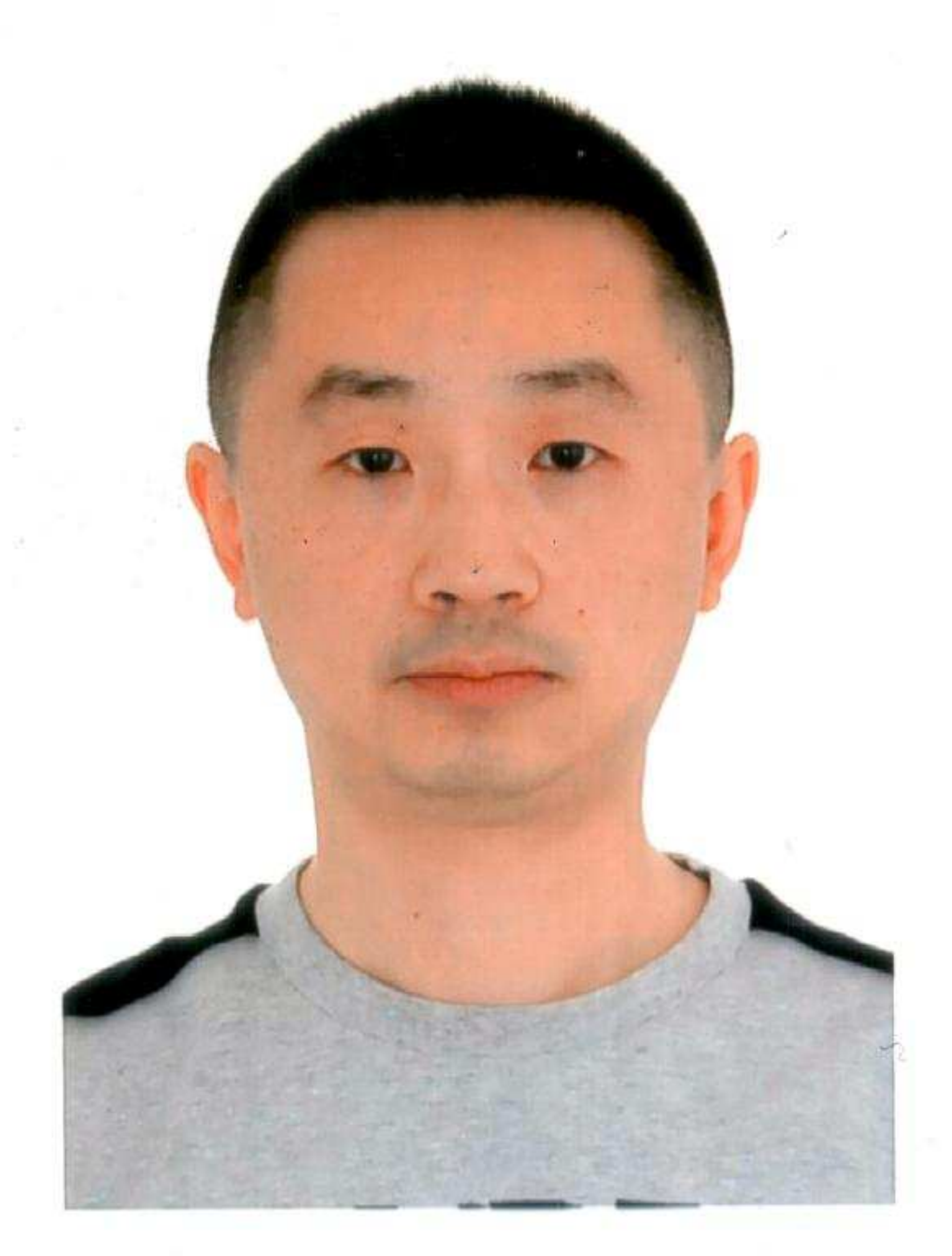}}]
\footnotesize \textbf{Ming Zhao}
received the M.Sc. and Ph.D. degrees in computer science from Central South University, Changsha, China, in 2003 and 2007, respectively. He is currently a Professor with the School of Computer Science and Engineering, Central South University. His main research focuses on wireless networks. He is also a Member of the China Computer Federation.
\end{IEEEbiography}

\begin{IEEEbiography}
[{\includegraphics[width=1in,height=1.25in,clip,keepaspectratio]{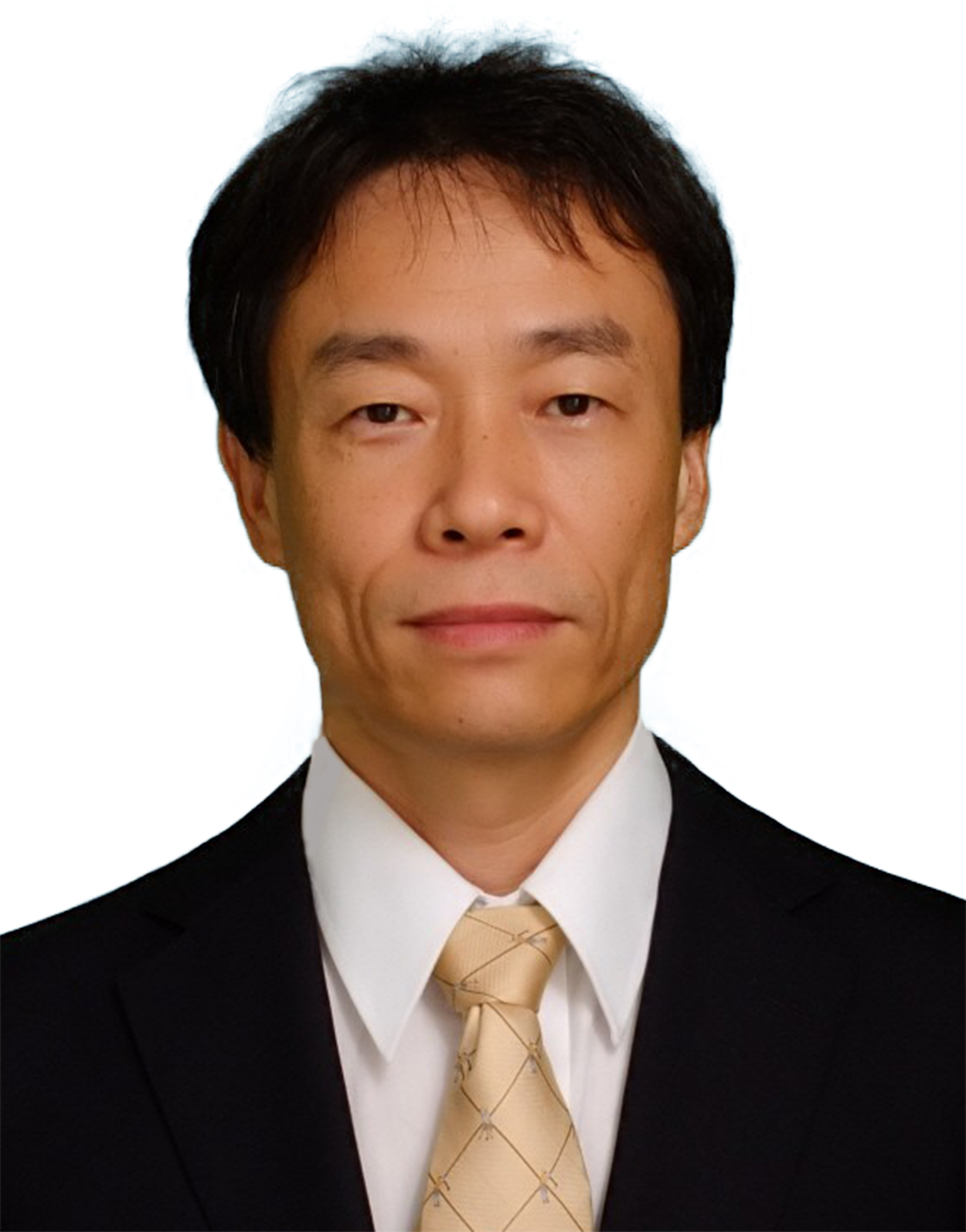}}]
\footnotesize \textbf{Nei Kato}
is a Full Professor and the Dean with the Graduate School of Information
Sciences, Tohoku University. He has researched on
computer networking, wireless mobile communications, satellite communications, ad hoc and sensor
and mesh networks, UAV networks, smart grid,
AI, IoT, Big Data, and pattern recognition. He has
published more than 500 papers in prestigious peerreviewed journals and conferences. He served as
the Vice-President (Member and Global Activities)
of IEEE Communications Society from 2018 to
2021, and the Editor-in-Chief of IEEE TRANSACTIONS ON VEHICULAR
TECHNOLOGY from 2017 to 2021. He is the Editor-in-Chief of IEEE
INTERNET OF THINGS JOURNAL. He is a Fellow of the Engineering Academy
of Japan and IEICE.

\end{IEEEbiography}

\vfill

\end{document}